\documentclass[aps,prb,twocolumn,groupedaddress]{revtex4-1}
\usepackage[version=3]{mhchem} 

\begin{document}

\title{Active Polarization Control with a Parity-Time Symmetric Plasmonic Resonator}

\author{Brian Baum}
\affiliation{Department of Materials Science and Engineering, Stanford University, Stanford, California 94305, USA}
\author{Mark Lawrence}
\affiliation{Department of Materials Science and Engineering, Stanford University, Stanford, California 94305, USA}
\author{David Barton III}
\affiliation{Department of Materials Science and Engineering, Stanford University, Stanford, California 94305, USA}
\author{Hadiseh Alaeian}
\affiliation{Department of Electrical Engineering, Stanford University, Stanford, California 94305, USA}
\altaffiliation{Current Address: 5. Physikalisches Institut, Universit{\"a}t Stuttgart, Pfaffenwaldring 57, 70569 Stuttgart, Germany}
\author{Jennifer Dionne}
\email{jdionne@stanford.edu}
\affiliation{Department of Materials Science and Engineering, Stanford University, Stanford, California 94305, USA}

\date{\today}

\begin{abstract}
Control of the polarization state of light is essential for many technologies, but is often limited by weak light-matter interactions that necessitate long device path lengths or significantly reduce the signal intensity.  Here, we investigate a nanoscale plasmonic aperture capable of modifying the polarization state of far-field transmitted light without loss in the probe signal. The aperture is a coaxial resonator consisting of a dielectric ring embedded within a metallic film; parity-time (PT) symmetric inclusions of loss and gain within the dielectric ring enable polarization control. Since the coaxial aperture enables near-thresholdless PT symmetry breaking, polarization control is achieved with realistic levels of loss and gain. Exploiting this sensitivity, we show that the aperture can function as a tunable waveplate, with the transmitted ellipticity of circularly polarized incident light changing continuously with the dissipation coefficient from pi/2 to 0 (i.e. linear polarization). Rotation of linearly polarized light with unity efficiency is also possible, with a continuously-tunable degree of rotation.  This compact, low-threshold, and reconfigurable polarizer may enable next-generation, high-efficiency displays, routers, modulators, and metasurfaces.

\end{abstract}

\pacs{}

\maketitle
\section{Introduction}
Polarization is a critical degree of freedom in a variety of optical components, including waveguides, filters, and detectors. As these components become increasingly small, efficient, and adaptive, polarizers must be designed with similar goals in mind. Compact and high-purity polarization has long been achieved using metal wire grid structures. Wires are relatively easy to fabricate on sub-micron scales, but the purity of the polarization comes at the expense of transmission; for example, the conversion of circularly polarized light to linearly polarized light reduces the transmitted intensity by 50\%. 

Improved efficiency can be achieved with polarization converters. For example, metasurfaces have enabled deeply subwavelength polarization conversion by controllably applying a phase delay or amplitude modulation to one polarization with respect to another\cite{Zhao2012,Meta-waveplates2016,Zhao2011,Guo:2013fq}. A variety of waveplate behaviors as well as spatial and spectral control have been achieved in both metallic and all-dielectric metasurfaces.\cite{Ellenbogen2012,Kildishev2013, Minovich2015, Zheludev2012,Yu2014, Li:2015ft, Kuznetsov:2016bv,Jiang:2016iy,Zhan:2016il,Jahani:2016fq}. More recently, these metasurfaces have also enabled reconfigurable optics, using liquid crystal layers\cite{Buchnev2015,Decker2013,Isic2015,Staude2015}, phase change materials\cite{Kats2012,Rensberg2016,Wang2015,Waters2015}, electrical modulation\cite{Capasso2013,Fallahi2012,Yao2014,Z.Fang2013, Leroux2009, Sun2013}, and mechanical modulation\cite{Pryce2010,Zheludev2016}. However, such transformations are usually binary, switching between two operating modes, and often come at the expense of increased device footprint.  

High-efficiency, reconfigurable polarizers may become active components for displays, routers, modulators, and processors. 
In this paper, we theoretically demonstrate how a continua of polarization states can be generated from an active, nanoscale plasmonic aperture. This aperture controls the amplitude of two orthogonal linear polarization states to create almost arbitrary output polarizations. Active control is achieved with a non-Hermitian, parity-time ($\mathcal{PT}$) symmetric configuration of loss and gain within a plasmonic coaxial aperture. Our design addresses three of the critical needs of future polarizers: (1) the device is nanoscale, (2) the device does not attenuate the transmitted intensity during the polarization conversion process, and (3) the device could be externally tuned by optical or electrical means to modify the output polarization. We show two test cases for polarization control: circular polarization conversion to linear polarization and rotation of linear polarizations. In both cases we focus on light scattered to the far-field in the direction normal to the surface. We also characterize the intensity and phase of a particular linear polarized component of this transmission.

$\mathcal{PT}$-symmetry is a relatively new tool added to the optical engineer's toolbox but has already enabled numerous photonic and plasmonic devices, ranging from ultra-sensitive sensors to efficient modulators, multiplexers, and directional lasers.\cite{LanNature2017, Feng:2012jj,Feng:2014gg,Baum:2015id,Ramezani:2010eb,Peng:2014kl,Lawrence2014, Alaeian:2014dj,Benisty:2011wq,Alaeian:2014eb} The balanced inclusion of gain and loss along an axis of symmetry (while maintaining a uniform real refractive index) describes the most general example of optical $\mathcal{PT}$-symmetry\cite{Guo:2009hd}. The magnitude of balanced gain and loss, i.e., the absolute value of the imaginary part of the refractive index, represents the deviation of a closed photonic system from Hermiticity. Accordingly, we refer to this value as the non-Hermiticity value, $\kappa$. 

Recently, the concept of PT symmetry has been extended to the polarization degree of freedom. Employing a periodic array of specially designed subwavelength resonators, or meta-atoms, distinct amplification or dissipation rates can be achieved for spatially co-located but orthogonally polarized light.\cite{Lawrence2014,yu2016acceleration, cerjan2017achieving, hassan2017dynamically}  These highly anisotropic systems have been shown to exhibit polarization phase transitions and polarization exceptional points, and therefore represent a novel mechanism for achieving efficient asymmetric polarization transformations within arbitrary basis representations. Previous investigations have however all relied upon coupling between different resonators or guided modes. In this case, the need to meet a structurally dependent gain threshold has limited experimental observations, especially in the optical regime. Precise coupling, absorption and dissipation coefficients also need to be maintained to get a desired output polarization.   

As the amount of loss and gain is increased, the eigenmodes of the system coalesce, becoming non-orthogonal and ultimately, degenerate at the exceptional point (EP). These EPs are remarkably sensitive to any system parameter variation, and so most $\mathcal{PT}$-symmetric devices are designed to operate at or beyond this EP based transition. The location of the EP in parameter space is controlled by the degree of non-Hermiticity and geometry. While a finite amount of loss and gain is generally required to reach an EP, recent designs based on modal degeneracy and temporal variation can enable thresholdless operation \cite{Ge:2014fd,Alaeian:2016hr, Tsampikos2017, Kante16}. These distributions help $\mathcal{PT}$-symmetric systems become realizable with physical values of gain and loss, and they are fully leveraged in the coaxial geometry under investigation in this work.

\begin{figure}
\includegraphics{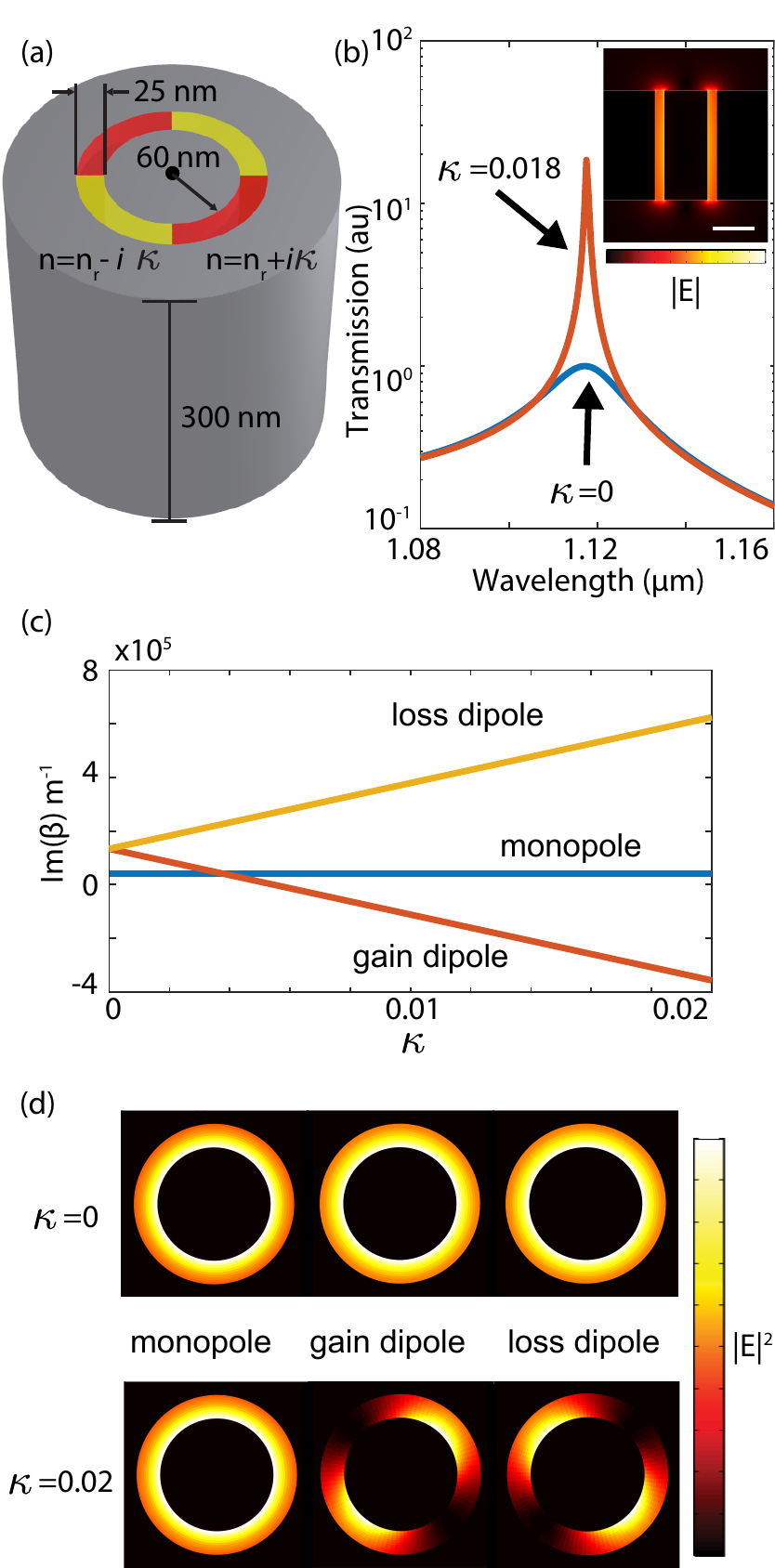}
\caption{\label{fig:transmission} (a) Schematic of the $\mathcal{PT}$-symmetric coaxial waveguide with alternating sections of the gain (yellow) and loss (red). (b) Transmission spectra of the 300-nm-long coaxial aperture when $\kappa=0$ and $\kappa=0.018$. Significant line-width narrowing and increased transmission intensity are present when $\kappa=0.018$. The inset is a cross section of the normalized electric field for $\kappa=0$ on resonance with a 120\,nm scale bar. (c) Dispersion of the imaginary wavevectors of an infinite $\mathcal{PT}$-symmetric coaxial waveguide with varying $\kappa$ at 1117\,nm. (d) Modal profiles of an infinite coaxial waveguide for $\kappa = 0$ and $0.02$.  Electric field intensity for modes labeled in part (c).  The loss and gain modes for $\kappa = 0.02$ are linearly polarized along the direction of high intensity. }
\label{fig:transmission}
\end{figure}

A schematic of the plasmonic coaxial aperture is shown in Figure~\ref{fig:transmission}(a). Note that similar resonator designs have been used to achieve extraordinary optical transmission, negative refractive indices, and low-threshold plasmonic lasers.\cite{Baida:2003gy,Baida:2006fi,Poujet:2007bg,Orbons:2007er,Burgos:2010ja,vandeHaar:2014eu,Saleh:2012em,Khajavikhan2012} As seen, the coaxial resonator consists of a $25$\,nm dielectric channel embedded within a $300$\,nm thick silver film. The core's radius is $60$\,nm, for a total coaxial cross section of $170$\,nm. The Ag is modeled with empirical data from Johnson and Christy.\cite{PhysRevB.6.4370}, to include realistic losses. The real part of the refractive index of the dielectric channel is $n=1.5$, while the imaginary part, $\pm\kappa$, is dynamically adjusted from $0$ to $0.0187$ during device operation. These values are achievable with traditional sources of gain media such as dopant dyes that could be introduced into a SiO$_{2}$ host\cite{berini2012surface, xiao2010loss,de2010amplification}. The distribution of positive or negative $\kappa$ is azimuthally defined as four alternating quads of gain and loss, producing two-fold mirror symmetry.
\section{Transmission of the $\mathcal{PT}$-symmetric coaxial resonator}
The transmission spectra of the finite coaxial aperture is highly dependent on the addition of gain and loss when illuminated with an incident plane wave. Figure~\ref{fig:transmission}(b) shows that the lowest order Fabry-Perot resonance of the aperture occurs at a wavelength of $1117$\,nm. The transmission is normalized to that of the $\kappa=0$ coaxial aperture and becomes ten times greater when $\kappa=0.018$ for linear polarized light aligned between the gain and loss sections. The inset of this figure shows a cross section of the field profile at $1117$\,nm, confirming this resonance as the lowest order Fabry-Perot mode. 

The sharp increase in transmission for even small amounts of gain and loss can be understood by considering the imaginary part of the wavevector for the mode supported by an infinitely-long coaxial waveguide. Figure~\ref{fig:transmission}(c) details the $\kappa$ dependence of the imaginary part of the three complex eigenvalues of the $\mathcal{PT}$-symmetric coaxial waveguide when excited with 1117\,nm light. The degeneracy of two modes is broken for $\kappa \neq 0$ producing amplifying or attenuating behavior (Im($\beta$) less than or greater than zero, respectively). In a previous work, we showed that these eigenvalues correspond to modes localized to the gain and loss sections in the coaxial waveguide, with near-linear polarization distributions.\cite{Alaeian:2016hr}

Because only the imaginary part of the channel refractive index is modulated, the resonance wavelength changes by less than a nanometer---a distinct advantage of this design over alternate phase-change material approaches. Although we report our findings for a specific wavelength with a specific geometry, future devices could easily be tailored across a wide spectral range by varying the dielectric channel thickness, the core diameter, and the aperture's length. Alternative dielectric fillers and metals could also tune the response of the resonator and would be particularly useful in regions of the electromagnetic spectrum where silver's response is not ideal.

\begin{figure}
\includegraphics{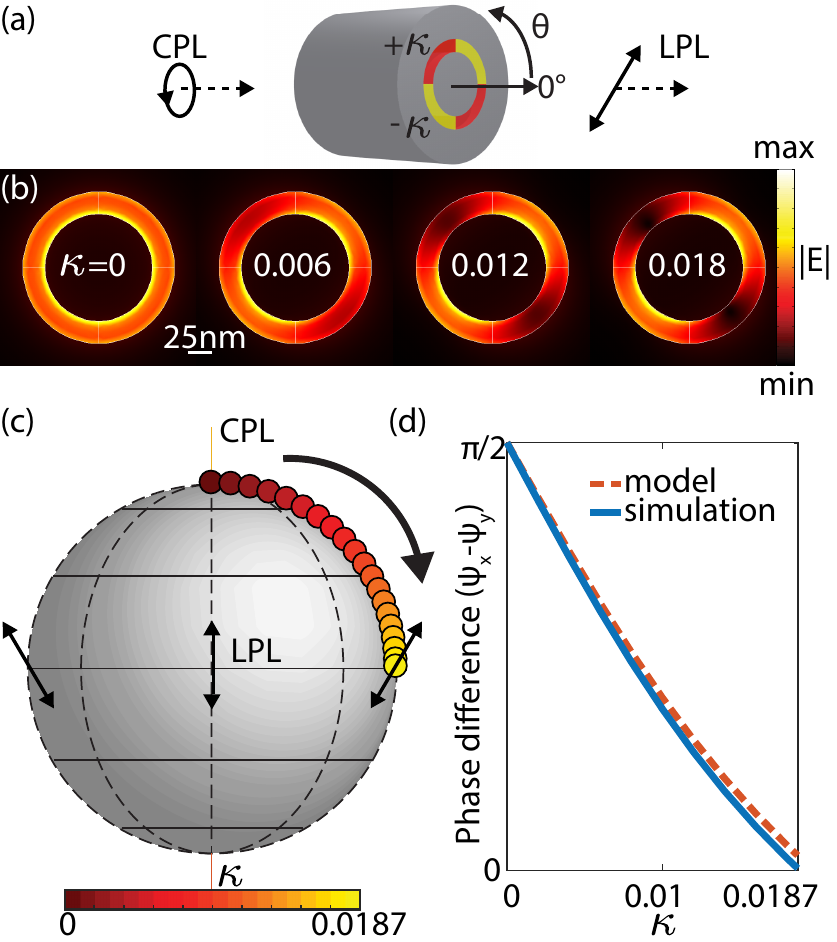}
\caption{\label{CPL} (a) Illustration of the illumination and transmission characteristics of the $\mathcal{PT}$-symmetric coaxial aperture showing CPL to LPL conversion. (b) Electric field intensity at the end of the aperture as a function of $\kappa$ when the aperture is illuminated with CPL. The intensity is azimuthally symmetric at $\kappa=0$ and becomes increasingly oriented towards the gain sections (45$^{\circ}$ and $225^{\circ}$) as $\kappa$ increases to $\kappa=0.018$. (c) Poincare sphere of the transmitted polarization state shows the far field polarization transitions from CPL to LPL. (d) Far-field phase difference between $E_x$ and $E_y$ shows good agreement between model and simulation for the transition from CPL to LPL.}
\label{fig:CPL}
\end{figure}

The total transmission for the $\mathcal{PT}$-symmetric coaxial aperture increases with $\kappa$, and results in distinct output polarization states in the far field. Our first example of polarization control is conversion from circularly polarized light (CPL) to linearly polarized light (LPL), schematically illustrated in Figure\,\ref{fig:CPL}(a). The near fields at the end of the aperture are displayed in Figure\,\ref{fig:CPL}(b) for select values of $\kappa$. The peak fields are found along the core of the coaxial channel for all values of $\kappa$, but the azimuthal distribution differs. Note that all fields are self-normalized, so the variations caused by increasing $\kappa$ are most evident in the decreased fields in the loss sections of the coaxial ring. When $\kappa=0.006$, the peak fields near the core interface are altered first, and the fields in the loss sections drop by roughly a factor of two compared to the peak fields in the gain sections. A node starts to appear in the middle of the loss sections for $\kappa=0.012$ and reaches near-minimum fields at $\kappa=0.018$. Throughout the full range of operation, the gain regions appear largely unchanged because of the self-normalization but experience well over a two-times amplification. The changes in the near-fields can be relevant for directional or locationally dependent near field coupling applications. To confirm the transmitted polarization state, we propagate the forward scattered light to the far-field, in the direction normal to the film surface, and analyze the transverse electric field components.

\section{Polarization conversion with circularly-polarized illumination}
Figure~\ref{fig:CPL}(c) shows the polarization state of the coaxial aperture's far-field transmission for a range of $\kappa$ when illuminated with CPL; we consider the fields normal to the coaxial aperture. The Poincare sphere is oriented such that the north and south poles correspond to CPL, while the equator is LPL. The polarization is marked with circles for $\kappa=0$ through $\kappa=0.0187$, as indicated by the colormap. Each point represents a $\kappa=0.01$ increment, aside from the last, which represents a 0.0087 increment. The points show a smooth progression from CPL, through varying degrees of ellipticity, to linear polarization. The polarization angle is approximately $45^\circ$, the angle corresponding to the center of the gain region. 

We next investigate the change in phase of the orthogonal polarizations. Figure~\ref{fig:CPL}(d) shows that the difference in phase between the electric field in $x$ and $y$ direction drops from $\frac{\pi}{2}$ to zero as the polarization transitions from circular to linear. The simulation shows a sublinear drop from a $\frac{\pi}{2}$ phase difference to a $0$ phase difference when $\kappa=0.0187$, the value at which the loss mode is diminished and the gain mode is sufficiently amplified.

To explain the rate of change of ellipticity as a function of $\kappa$, we develop a basic model that captures the anisotropic amplitude modification. Briefly, we consider two orthogonal linear polarizations, one aligned to the loss sections of the coaxial resonator, and one aligned to the gain sections. Note that these states form a basis from which all polarizations can be built through variations in phase or amplitude. The loss-aligned polarization ($-45^\circ$) experiences enhanced absorption as $\kappa$ is increased, and the gain-aligned polarization ($45^\circ$) experiences less absorption or even amplification. Variations in the orthogonal polarizations allow the coaxial aperture to effectively pull an input polarization state towards a linear polarization state aligned with the gain sections. Since the aperture modifies only the intensities of the orthogonal polarizations, we can model the transmission of the two linear polarizations as Lorentzian oscillators and consider their peak transmission on resonance as:

\begin{align} 
   T_{peak}=\frac{a}{b \pm \kappa},
\end{align}

where $a$ and $b$ relate to the scattering cross section and lifetime of the resonance at $\kappa=0$. By fitting the transmission dependence as a function of $\kappa$ for the two orthogonal polarizations, we determine $a=6.3967e^{-5}$ and $b=0.0197$. This basic model well-described the phase-difference behavior in our coaxial polarizer, seen in Figure 2(d). 
\section{Polarization conversion with Linearly-polarized illumination}
\begin{figure}
\includegraphics{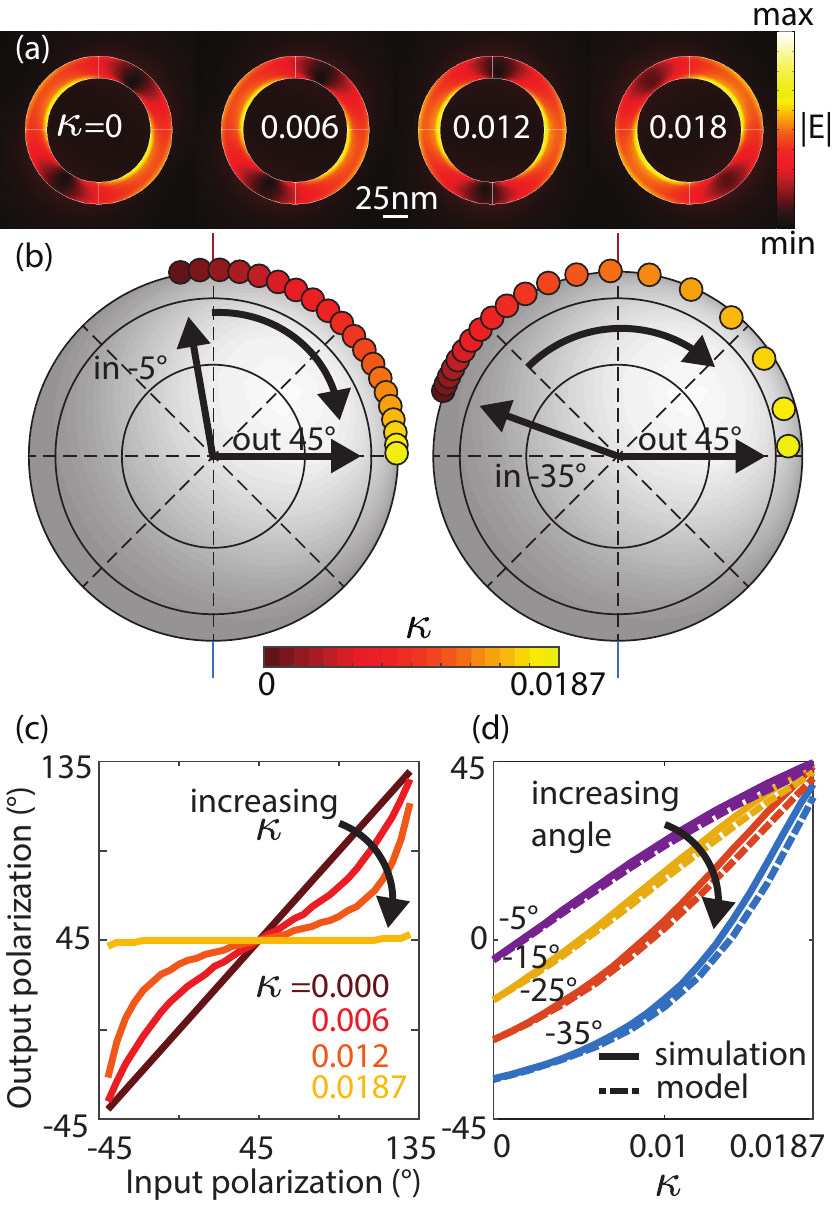}
\caption{\label{LPL} (a) Electric field intensity at the end of the aperture when the aperture is illuminated with LPL aligned to $-35^\circ$. The field intensity rotates around the coaxial structure.  Each field is self-normalized, with the intensity for $\kappa=0.0187$ being approximately an order of magnitude lower than that in figure 2(a). The output intensity is still greater than the passive ($\kappa=0$) case. (b) Poincare spheres are oriented so that the perimeter of the projection represents linear polarization. The polarization state moves along the perimeter. Input polarizations of $-5^\circ$ and $-35^\circ$ rotate with increasing $\kappa$ to the gain angle of $45^\circ$. (c) Output polarization angle as a function of the input polarization and $\kappa$, shows polarization is pulled towards the gain axis ($45^\circ$) as $\kappa$ is increased to 0.0187. (e) Simulation and model of the output polarization as a function of $\kappa$ showing $-5^\circ$,  $-15^\circ$, $-25^\circ$, and $-35^\circ$ input polarizations rotated to $45^\circ$.}
\label{fig:LPL}
\end{figure}

Selective amplification and absorption can also be used to achieve continuous polarization rotation of linearly-polarized light. Here, the non-Hermiticity parameter serves to adjust the relative amplitudes of the orthogonal field components. The electric field intensity at the end of the coaxial aperture is shown in Figure~\ref{fig:LPL}(a) when the structure is illuminated with $-35^\circ$ linearly polarized light for four values of $\kappa$. $-35^\circ$ corresponds to $10^\circ$ away from the loss axis, and so the electric fields are localized to the loss sections while field nodes exist in the gain sections. As $\kappa$ is increased, the electromagnetic hot spots and nodes rotate counter-clockwise around the dielectric ring, appearing to almost straddle the divide between the gain and loss sections when $\kappa=0.012$. Beyond $\kappa=0.012$, the rotation increment per $\kappa$ increases; by $\kappa=0.018$, we see that the electromagnetic hot spots have aligned with the gain sections, and the weaker nodes are aligned with the loss sections. The slight elevation of the field minima suggests that the polarization has taken on some minor ellipticity for this full $80^\circ$ rotation. For input polarizations beyond $85^\circ$ offset from the gain-angle, we see the coaxial aperture functions as a polarization filter and increasingly absorbs light for higher values of $\kappa$. LPL inputs between $-40^\circ$ and $-50^\circ$ will therefore lose some intensity when passed through the coaxial aperture. 

The progression of the normal far field polarization rotation is illustrated for two input linear polarizations, $-5^\circ$ and $-35^\circ$, in Figure~\ref{fig:LPL}(b). The Poincare spheres are rotated such that the perimeter corresponds to linear polarized outputs (equator). For both inputs, we see that for $\kappa=0.0187$ the polarization is pulled to $45^\circ$, the angle corresponding to the gain sections. When the input is $-5^\circ$, the points are relatively equally spaced and lie directly on the perimeter. In fact, for the final $45^\circ$ output, the ellipticity ratio between the gain aligned polarization and the loss aligned polarization (major axis and minor axis) is over 550. Conversely, when the input polarization is $-35^\circ$, the spacing between the equal steps in $\kappa$ appears nonlinear, and for the maximum degree of rotation the point lies slightly off the perimeter, indicating some ellipticity. For this maximum range of $80^\circ$ rotation, the elliptical contrast ratio between the gain and loss axis (major axis and minor axis) is approximately 25.

A complete range of input polarizations and their resulting far-field output polarizations is shown in Figure~\ref{fig:LPL}(c) for four select values of $\kappa$. All input polarizations which are offset less than $80^\circ$ are fully rotated to $45^\circ$. We see the polarizations are pulled down to the gain axis at $45^\circ$ with nonlinear rates that vary both as a function of input polarization and $\kappa$. The nonlinearity of the polarization rotation is explored more thoroughly in Figure~\ref{fig:LPL}(d). For $-5^\circ$ and $-15^\circ$ inputs, we see roughly linear polarization rotation as a function of $\kappa$ in both the simulation and the model. As the total distance of rotation is increased, as in the case of $-25^\circ$ and $-35^\circ$ inputs, we see the sensitivity of the output polarization to $\kappa$ increases with $\kappa$ in both the simulation and the model. This nonlinearity arises from the inverse relationship between the transmitted fields and $\kappa$. The model shows good agreement with the simulation for all cases, often overlapping with one another, and differs only at the highest values of $\kappa$. What little difference between the simulation and model we do see in Figure~\ref{fig:LPL}(d) and Figure~\ref{fig:CPL}(d) can be attributed to a slight resonance wavelength mismatch  ($<1$\,nm) of the gain-aligned and loss-aligned transmission. The effect is more pronounced at high values of $\kappa$, as the line width of the gain-aligned mode decreases substantially.

\begin{figure}
\includegraphics{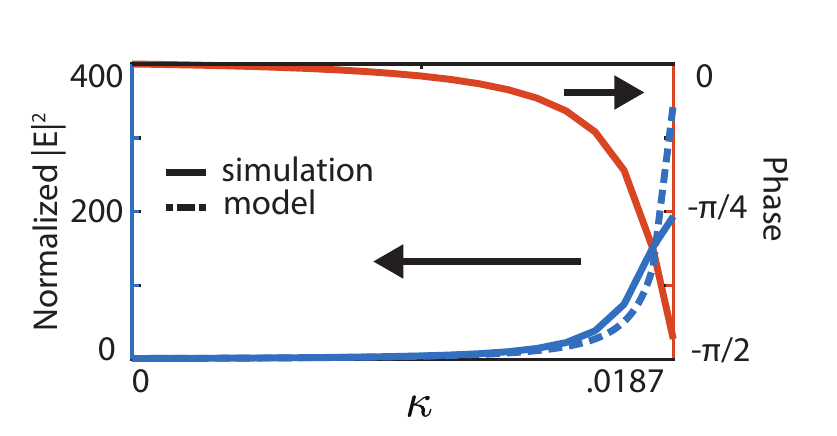}
\caption{\label{phaseamp} Phase and amplitude as a function of $\kappa$ for the polarization aligned to the gain axis. The model transmission intensity is plotted as a dotted line.}
\label{fig:phaseamp}
\end{figure}

Although we motivate the $\mathcal{PT}$-symmetric coaxial aperture as a means of altering the polarization, if the coaxial aperture is paired with a polarizer it may also function as an active amplitude and phase modulator. In Figure~\ref{fig:phaseamp}, we show that if the output of the aperture is filtered to be aligned with the gain-axis at $45^\circ$ with a separate polarizer, the total far-field normal intensity varies by roughly 200 times regardless of input polarization condition. A similar trend is observed in the basic model described above, thought the model overestimates the intensity, as it does not include the slight peak shift. The peak shift can also be used to modulate the phase of transmitted light, as evidenced by the nearly $\frac{\pi}{2}$ phase shift.
\section{Summary and Outlook}
We have theoretically demonstrated a nanoscale system for active polarization control using parity-time symmetric inclusions of gain and loss within a plasmonic aperture. The aperture is capable of transforming circularly polarized light to linearly polarized light by increasing the imaginary part of the refractive index in the gain and loss dielectric materials. Linear polarization rotation is also achieved through the same mechanism. Active polarization control could be experimentally realized with realistic materials by simply changing the optical or electronic pump conditions of this medium.  Further, gain saturation in this structure could enable a variety of thresholdless nonlinear phenomena, including wireless power transfer\cite{assawaworrarit2017robust}, on-chip isolation\cite{chang2014parity}, and broadband nonreciprocity\cite{barton2018broadband}. Both in the linear and nonlinear regime, these nanoscale coaxial resonators could be relevant for polarization control when coupled to emitters; for thin-film polarization filtering in display technologies; and for phase, polarization, or amplitude controlling metasurfaces.

\section*{Acknowledgements}
The authors appreciate all Dionne group members for their insightful feedback on the work. Funding from a Presidential Early Career Award administered through the Air Force Office of Scientific Research (grant no. FA9550-15-1-0006) is also gratefully acknowledged.
\appendix

\section{COMSOL simulation details}
The simulations were set up in three electromagnetic waves frequency domain (EWFD) sessions in the wave optics module. The simulation volume was 1200\,nm by 1200\,nm wide with a free space height of 400\,nm above and below the metal. Two EWFDs were used to create the plane wave excitation field for the coaxial aperture, and each possessed a single port with orthogonally polarized linear plane wave inputs. The two excitation EWFDs were bound with the appropriate perfect electric and perfect magnetic boundary conditions to generate their respectively polarized plane wave background fields in the free space volume before the coaxial aperture. The third EWFD was the scattering transmission environment. This final simulation was fed the fields of the first two simulations as a background field. The scattering simulation was surrounded in a scattering boundary conditions and perfectly matched layers to attenuate the scattered fields. 

The far-field monitor was placed on the opposite side of the metal aperture from the excitation ports 50\,nm away from the surface. The monitor was a half-sphere with a radius of 200\,nm. Far fields were determined using COMSOL's far-field domain solver, which uses the Stratton-Chu method. We found the monitor and solver produced little variation in the far-fields propagated normal to the surface across a variety of monitor sizes and separations.

The whole simulation volume was meshed with tetragonal cells, with a maximum size of 8\,nm in the coaxial channel, 30\,nm in the metal, and 100\,nm in free space. The PML was 150\,nm thick with five layers. We saw no variation in the reported values when the thickness of the PML regions was increased.
\section{Derivation of Equation 1}
We can model the optical response of the $\mathcal{PT}$ coaxial aperture as two perpendicular dipole antennas, one aligned with the center of the gain segments and the other aligned with the center of the loss segments. Such a dipole with a passive dielectric inclusion can be described by a Lorentzian function,
\begin{equation}
p_i=\frac{a_i}{\omega-\omega_i+ib_i}E_i
\end{equation}
where $p_i$, $E_i$, $a_i$, $b_i$, and $\omega_i$ represent the dipole emission, incident field strength, dipole strength, dissipation rate (consisting of both radiative and nonradiative mechanisms) and resonant frequency in the ith direction, respectively. With small loss and gain added to the dielectric inclusion we expect only a linear modification to the dissipation coefficient, $b_i\rightarrow b_i\pm \kappa$.On resonance, $\omega_g=\omega_l=\omega$ and the associated transmission can then be written as  
\begin{equation}
T_l=\frac{a}{b+\kappa}E_l
\end{equation}
and
\begin{equation}
T_g=\frac{a}{b-\kappa}E_g
\end{equation}

\bibliography{polconv}

\begin{thebibliography}{63}%
\makeatletter
\providecommand \@ifxundefined [1]{%
 \@ifx{#1\undefined}
}%
\providecommand \@ifnum [1]{%
 \ifnum #1\expandafter \@firstoftwo
 \else \expandafter \@secondoftwo
 \fi
}%
\providecommand \@ifx [1]{%
 \ifx #1\expandafter \@firstoftwo
 \else \expandafter \@secondoftwo
 \fi
}%
\providecommand \natexlab [1]{#1}%
\providecommand \enquote  [1]{``#1''}%
\providecommand \bibnamefont  [1]{#1}%
\providecommand \bibfnamefont [1]{#1}%
\providecommand \citenamefont [1]{#1}%
\providecommand \href@noop [0]{\@secondoftwo}%
\providecommand \href [0]{\begingroup \@sanitize@url \@href}%
\providecommand \@href[1]{\@@startlink{#1}\@@href}%
\providecommand \@@href[1]{\endgroup#1\@@endlink}%
\providecommand \@sanitize@url [0]{\catcode `\\12\catcode `\$12\catcode
  `\&12\catcode `\#12\catcode `\^12\catcode `\_12\catcode `\%12\relax}%
\providecommand \@@startlink[1]{}%
\providecommand \@@endlink[0]{}%
\providecommand \url  [0]{\begingroup\@sanitize@url \@url }%
\providecommand \@url [1]{\endgroup\@href {#1}{\urlprefix }}%
\providecommand \urlprefix  [0]{URL }%
\providecommand \Eprint [0]{\href }%
\providecommand \doibase [0]{http://dx.doi.org/}%
\providecommand \selectlanguage [0]{\@gobble}%
\providecommand \bibinfo  [0]{\@secondoftwo}%
\providecommand \bibfield  [0]{\@secondoftwo}%
\providecommand \translation [1]{[#1]}%
\providecommand \BibitemOpen [0]{}%
\providecommand \bibitemStop [0]{}%
\providecommand \bibitemNoStop [0]{.\EOS\space}%
\providecommand \EOS [0]{\spacefactor3000\relax}%
\providecommand \BibitemShut  [1]{\csname bibitem#1\endcsname}%
\let\auto@bib@innerbib\@empty
\bibitem [{\citenamefont {Zhao}\ \emph {et~al.}(2012)\citenamefont {Zhao},
  \citenamefont {Belkin},\ and\ \citenamefont {Al{\`{u}}}}]{Zhao2012}%
  \BibitemOpen
  \bibfield  {author} {\bibinfo {author} {\bibfnamefont {Y.}~\bibnamefont
  {Zhao}}, \bibinfo {author} {\bibfnamefont {M.}~\bibnamefont {Belkin}}, \ and\
  \bibinfo {author} {\bibfnamefont {A.}~\bibnamefont {Al{\`{u}}}},\ }\href
  {\doibase 10.1038/ncomms1877} {\bibfield  {journal} {\bibinfo  {journal}
  {Nature Communications}\ }\textbf {\bibinfo {volume} {3}},\ \bibinfo {pages}
  {870} (\bibinfo {year} {2012})}\BibitemShut {NoStop}%
\bibitem [{\citenamefont {Zhao}\ and\ \citenamefont
  {Al{\`{u}}}(2013)}]{Meta-waveplates2016}%
  \BibitemOpen
  \bibfield  {author} {\bibinfo {author} {\bibfnamefont {Y.}~\bibnamefont
  {Zhao}}\ and\ \bibinfo {author} {\bibfnamefont {A.}~\bibnamefont
  {Al{\`{u}}}},\ }\href {\doibase 10.1021/nl304392b} {\bibfield  {journal}
  {\bibinfo  {journal} {Nano Letters}\ }\textbf {\bibinfo {volume} {13}},\
  \bibinfo {pages} {1086} (\bibinfo {year} {2013})}\BibitemShut {NoStop}%
\bibitem [{\citenamefont {Zhao}\ and\ \citenamefont
  {Al{\`{u}}}(2011)}]{Zhao2011}%
  \BibitemOpen
  \bibfield  {author} {\bibinfo {author} {\bibfnamefont {Y.}~\bibnamefont
  {Zhao}}\ and\ \bibinfo {author} {\bibfnamefont {A.}~\bibnamefont
  {Al{\`{u}}}},\ }\href {\doibase 10.1103/PhysRevB.84.205428} {\bibfield
  {journal} {\bibinfo  {journal} {Physical Review B - Condensed Matter and
  Materials Physics}\ }\textbf {\bibinfo {volume} {84}},\ \bibinfo {pages} {1}
  (\bibinfo {year} {2011})}\BibitemShut {NoStop}%
\bibitem [{\citenamefont {Guo}\ \emph {et~al.}(2013)\citenamefont {Guo},
  \citenamefont {Yang}, \citenamefont {Li}, \citenamefont {Guo}, \citenamefont
  {Cui}, \citenamefont {Kang},\ and\ \citenamefont {Chen}}]{Guo:2013fq}%
  \BibitemOpen
  \bibfield  {author} {\bibinfo {author} {\bibfnamefont {Q.-H.}\ \bibnamefont
  {Guo}}, \bibinfo {author} {\bibfnamefont {M.}~\bibnamefont {Yang}}, \bibinfo
  {author} {\bibfnamefont {T.-F.}\ \bibnamefont {Li}}, \bibinfo {author}
  {\bibfnamefont {T.-J.}\ \bibnamefont {Guo}}, \bibinfo {author} {\bibfnamefont
  {H.-X.}\ \bibnamefont {Cui}}, \bibinfo {author} {\bibfnamefont
  {M.}~\bibnamefont {Kang}}, \ and\ \bibinfo {author} {\bibfnamefont
  {J.}~\bibnamefont {Chen}},\ }\href@noop {} {\bibfield  {journal} {\bibinfo
  {journal} {Applied Physics Letters}\ }\textbf {\bibinfo {volume} {102}},\
  \bibinfo {pages} {211906} (\bibinfo {year} {2013})}\BibitemShut {NoStop}%
\bibitem [{\citenamefont {Ellenbogen}\ \emph {et~al.}(2012)\citenamefont
  {Ellenbogen}, \citenamefont {Seo},\ and\ \citenamefont
  {Crozier}}]{Ellenbogen2012}%
  \BibitemOpen
  \bibfield  {author} {\bibinfo {author} {\bibfnamefont {T.}~\bibnamefont
  {Ellenbogen}}, \bibinfo {author} {\bibfnamefont {K.}~\bibnamefont {Seo}}, \
  and\ \bibinfo {author} {\bibfnamefont {K.~B.}\ \bibnamefont {Crozier}},\
  }\href {\doibase 10.1021/nl204257g} {\bibfield  {journal} {\bibinfo
  {journal} {Nano Letters}\ }\textbf {\bibinfo {volume} {12}},\ \bibinfo
  {pages} {1026} (\bibinfo {year} {2012})}\BibitemShut {NoStop}%
\bibitem [{\citenamefont {Kildishev}\ \emph {et~al.}(2013)\citenamefont
  {Kildishev}, \citenamefont {Boltasseva},\ and\ \citenamefont
  {Shalaev}}]{Kildishev2013}%
  \BibitemOpen
  \bibfield  {author} {\bibinfo {author} {\bibfnamefont {A.~V.}\ \bibnamefont
  {Kildishev}}, \bibinfo {author} {\bibfnamefont {A.}~\bibnamefont
  {Boltasseva}}, \ and\ \bibinfo {author} {\bibfnamefont {V.~M.}\ \bibnamefont
  {Shalaev}},\ }\href
  {http://science.sciencemag.org/content/339/6125/1232009.abstract} {\bibfield
  {journal} {\bibinfo  {journal} {Science}\ }\textbf {\bibinfo {volume} {339}}
  (\bibinfo {year} {2013})}\BibitemShut {NoStop}%
\bibitem [{\citenamefont {Minovich}\ \emph {et~al.}(2015)\citenamefont
  {Minovich}, \citenamefont {Miroshnichenko}, \citenamefont {Bykov},
  \citenamefont {Murzina}, \citenamefont {Neshev},\ and\ \citenamefont
  {Kivshar}}]{Minovich2015}%
  \BibitemOpen
  \bibfield  {author} {\bibinfo {author} {\bibfnamefont {A.~E.}\ \bibnamefont
  {Minovich}}, \bibinfo {author} {\bibfnamefont {A.~E.}\ \bibnamefont
  {Miroshnichenko}}, \bibinfo {author} {\bibfnamefont {A.~Y.}\ \bibnamefont
  {Bykov}}, \bibinfo {author} {\bibfnamefont {T.~V.}\ \bibnamefont {Murzina}},
  \bibinfo {author} {\bibfnamefont {D.~N.}\ \bibnamefont {Neshev}}, \ and\
  \bibinfo {author} {\bibfnamefont {Y.~S.}\ \bibnamefont {Kivshar}},\ }\href
  {\doibase 10.1002/lpor.201400402} {\bibfield  {journal} {\bibinfo  {journal}
  {Laser and Photonics Reviews}\ }\textbf {\bibinfo {volume} {9}},\ \bibinfo
  {pages} {195} (\bibinfo {year} {2015})}\BibitemShut {NoStop}%
\bibitem [{\citenamefont {Zheludev}\ and\ \citenamefont
  {Kivshar}(2012)}]{Zheludev2012}%
  \BibitemOpen
  \bibfield  {author} {\bibinfo {author} {\bibfnamefont {N.~I.}\ \bibnamefont
  {Zheludev}}\ and\ \bibinfo {author} {\bibfnamefont {Y.~S.}\ \bibnamefont
  {Kivshar}},\ }\href {\doibase 10.1038/nmat3431} {\bibfield  {journal}
  {\bibinfo  {journal} {Nature Materials}\ }\textbf {\bibinfo {volume} {11}},\
  \bibinfo {pages} {917} (\bibinfo {year} {2012})},\ \Eprint
  {http://arxiv.org/abs/1232009} {arXiv:1232009} \BibitemShut {NoStop}%
\bibitem [{\citenamefont {Yu}\ and\ \citenamefont {Capasso}(2014)}]{Yu2014}%
  \BibitemOpen
  \bibfield  {author} {\bibinfo {author} {\bibfnamefont {N.}~\bibnamefont
  {Yu}}\ and\ \bibinfo {author} {\bibfnamefont {F.}~\bibnamefont {Capasso}},\
  }\href {\doibase 10.1038/nmat3839} {\bibfield  {journal} {\bibinfo  {journal}
  {Nature Materials}\ }\textbf {\bibinfo {volume} {13}},\ \bibinfo {pages}
  {139} (\bibinfo {year} {2014})}\BibitemShut {NoStop}%
\bibitem [{\citenamefont {Li}\ \emph {et~al.}(2015)\citenamefont {Li},
  \citenamefont {Liu}, \citenamefont {Cheng}, \citenamefont {Chen},\ and\
  \citenamefont {Tian}}]{Li:2015ft}%
  \BibitemOpen
  \bibfield  {author} {\bibinfo {author} {\bibfnamefont {Z.}~\bibnamefont
  {Li}}, \bibinfo {author} {\bibfnamefont {W.}~\bibnamefont {Liu}}, \bibinfo
  {author} {\bibfnamefont {H.}~\bibnamefont {Cheng}}, \bibinfo {author}
  {\bibfnamefont {S.}~\bibnamefont {Chen}}, \ and\ \bibinfo {author}
  {\bibfnamefont {J.}~\bibnamefont {Tian}},\ }\href@noop {} {\bibfield
  {journal} {\bibinfo  {journal} {Scientific Reports}\ }\textbf {\bibinfo
  {volume} {5}},\ \bibinfo {pages} {18106} (\bibinfo {year}
  {2015})}\BibitemShut {NoStop}%
\bibitem [{\citenamefont {Kuznetsov}\ \emph {et~al.}(2016)\citenamefont
  {Kuznetsov}, \citenamefont {Miroshnichenko}, \citenamefont {Brongersma},
  \citenamefont {Kivshar},\ and\ \citenamefont
  {Luk'yanchuk}}]{Kuznetsov:2016bv}%
  \BibitemOpen
  \bibfield  {author} {\bibinfo {author} {\bibfnamefont {A.~I.}\ \bibnamefont
  {Kuznetsov}}, \bibinfo {author} {\bibfnamefont {A.~E.}\ \bibnamefont
  {Miroshnichenko}}, \bibinfo {author} {\bibfnamefont {M.~L.}\ \bibnamefont
  {Brongersma}}, \bibinfo {author} {\bibfnamefont {Y.~S.}\ \bibnamefont
  {Kivshar}}, \ and\ \bibinfo {author} {\bibfnamefont {B.}~\bibnamefont
  {Luk'yanchuk}},\ }\href@noop {} {\bibfield  {journal} {\bibinfo  {journal}
  {Science}\ }\textbf {\bibinfo {volume} {354}},\ \bibinfo {pages} {aag2472}
  (\bibinfo {year} {2016})}\BibitemShut {NoStop}%
\bibitem [{\citenamefont {Jiang}\ \emph {et~al.}(2016)\citenamefont {Jiang},
  \citenamefont {Zhao},\ and\ \citenamefont {Jiang}}]{Jiang:2016iy}%
  \BibitemOpen
  \bibfield  {author} {\bibinfo {author} {\bibfnamefont {H.}~\bibnamefont
  {Jiang}}, \bibinfo {author} {\bibfnamefont {W.}~\bibnamefont {Zhao}}, \ and\
  \bibinfo {author} {\bibfnamefont {Y.}~\bibnamefont {Jiang}},\ }\href@noop {}
  {\bibfield  {journal} {\bibinfo  {journal} {Optics Express}\ }\textbf
  {\bibinfo {volume} {24}},\ \bibinfo {pages} {17738} (\bibinfo {year}
  {2016})}\BibitemShut {NoStop}%
\bibitem [{\citenamefont {Zhan}\ \emph {et~al.}(2016)\citenamefont {Zhan},
  \citenamefont {Colburn}, \citenamefont {Trivedi}, \citenamefont {Fryett},
  \citenamefont {Dodson},\ and\ \citenamefont {Majumdar}}]{Zhan:2016il}%
  \BibitemOpen
  \bibfield  {author} {\bibinfo {author} {\bibfnamefont {A.}~\bibnamefont
  {Zhan}}, \bibinfo {author} {\bibfnamefont {S.}~\bibnamefont {Colburn}},
  \bibinfo {author} {\bibfnamefont {R.}~\bibnamefont {Trivedi}}, \bibinfo
  {author} {\bibfnamefont {T.~K.}\ \bibnamefont {Fryett}}, \bibinfo {author}
  {\bibfnamefont {C.~M.}\ \bibnamefont {Dodson}}, \ and\ \bibinfo {author}
  {\bibfnamefont {A.}~\bibnamefont {Majumdar}},\ }\href@noop {} {\bibfield
  {journal} {\bibinfo  {journal} {ACS Photonics}\ }\textbf {\bibinfo {volume}
  {3}},\ \bibinfo {pages} {209} (\bibinfo {year} {2016})}\BibitemShut {NoStop}%
\bibitem [{\citenamefont {Jahani}\ and\ \citenamefont
  {Jacob}(2016)}]{Jahani:2016fq}%
  \BibitemOpen
  \bibfield  {author} {\bibinfo {author} {\bibfnamefont {S.}~\bibnamefont
  {Jahani}}\ and\ \bibinfo {author} {\bibfnamefont {Z.}~\bibnamefont {Jacob}},\
  }\href@noop {} {\bibfield  {journal} {\bibinfo  {journal} {Nature
  Nanotechnology}\ }\textbf {\bibinfo {volume} {11}},\ \bibinfo {pages} {23}
  (\bibinfo {year} {2016})}\BibitemShut {NoStop}%
\bibitem [{\citenamefont {Buchnev}\ \emph {et~al.}(2015)\citenamefont
  {Buchnev}, \citenamefont {Podoliak}, \citenamefont {Kaczmarek}, \citenamefont
  {Zheludev},\ and\ \citenamefont {Fedotov}}]{Buchnev2015}%
  \BibitemOpen
  \bibfield  {author} {\bibinfo {author} {\bibfnamefont {O.}~\bibnamefont
  {Buchnev}}, \bibinfo {author} {\bibfnamefont {N.}~\bibnamefont {Podoliak}},
  \bibinfo {author} {\bibfnamefont {M.}~\bibnamefont {Kaczmarek}}, \bibinfo
  {author} {\bibfnamefont {N.~I.}\ \bibnamefont {Zheludev}}, \ and\ \bibinfo
  {author} {\bibfnamefont {V.~A.}\ \bibnamefont {Fedotov}},\ }\href {\doibase
  10.1002/adom.201400494} {\bibfield  {journal} {\bibinfo  {journal} {Advanced
  Optical Materials}\ }\textbf {\bibinfo {volume} {3}},\ \bibinfo {pages} {674}
  (\bibinfo {year} {2015})}\BibitemShut {NoStop}%
\bibitem [{\citenamefont {Decker}\ \emph {et~al.}(2013)\citenamefont {Decker},
  \citenamefont {Kremers}, \citenamefont {Minovich}, \citenamefont {Staude},
  \citenamefont {Miroshnichenko}, \citenamefont {Chigrin}, \citenamefont
  {Neshev}, \citenamefont {Jagadish},\ and\ \citenamefont
  {Kivshar}}]{Decker2013}%
  \BibitemOpen
  \bibfield  {author} {\bibinfo {author} {\bibfnamefont {M.}~\bibnamefont
  {Decker}}, \bibinfo {author} {\bibfnamefont {C.}~\bibnamefont {Kremers}},
  \bibinfo {author} {\bibfnamefont {A.}~\bibnamefont {Minovich}}, \bibinfo
  {author} {\bibfnamefont {I.}~\bibnamefont {Staude}}, \bibinfo {author}
  {\bibfnamefont {A.~E.}\ \bibnamefont {Miroshnichenko}}, \bibinfo {author}
  {\bibfnamefont {D.}~\bibnamefont {Chigrin}}, \bibinfo {author} {\bibfnamefont
  {D.~N.}\ \bibnamefont {Neshev}}, \bibinfo {author} {\bibfnamefont
  {C.}~\bibnamefont {Jagadish}}, \ and\ \bibinfo {author} {\bibfnamefont
  {Y.~S.}\ \bibnamefont {Kivshar}},\ }\href {\doibase 10.1364/OE.21.008879}
  {\bibfield  {journal} {\bibinfo  {journal} {Optics Express}\ }\textbf
  {\bibinfo {volume} {21}},\ \bibinfo {pages} {8879} (\bibinfo {year}
  {2013})},\ \Eprint {http://arxiv.org/abs/1302.4484} {arXiv:1302.4484}
  \BibitemShut {NoStop}%
\bibitem [{\citenamefont {Isi{\'{c}}}\ \emph {et~al.}(2015)\citenamefont
  {Isi{\'{c}}}, \citenamefont {Vasi{\'{c}}}, \citenamefont {Zografopoulos},
  \citenamefont {Beccherelli},\ and\ \citenamefont {Gaji{\'{c}}}}]{Isic2015}%
  \BibitemOpen
  \bibfield  {author} {\bibinfo {author} {\bibfnamefont {G.}~\bibnamefont
  {Isi{\'{c}}}}, \bibinfo {author} {\bibfnamefont {B.}~\bibnamefont
  {Vasi{\'{c}}}}, \bibinfo {author} {\bibfnamefont {D.~C.}\ \bibnamefont
  {Zografopoulos}}, \bibinfo {author} {\bibfnamefont {R.}~\bibnamefont
  {Beccherelli}}, \ and\ \bibinfo {author} {\bibfnamefont {R.}~\bibnamefont
  {Gaji{\'{c}}}},\ }\href {\doibase 10.1103/PhysRevApplied.3.064007} {\bibfield
   {journal} {\bibinfo  {journal} {Physical Review Applied}\ }\textbf {\bibinfo
  {volume} {3}},\ \bibinfo {pages} {1} (\bibinfo {year} {2015})}\BibitemShut
  {NoStop}%
\bibitem [{\citenamefont {Staude}\ \emph {et~al.}(2015)\citenamefont {Staude},
  \citenamefont {Decker}, \citenamefont {Rusak}, \citenamefont {Neshev},\ and\
  \citenamefont {Brener}}]{Staude2015}%
  \BibitemOpen
  \bibfield  {author} {\bibinfo {author} {\bibfnamefont {I.}~\bibnamefont
  {Staude}}, \bibinfo {author} {\bibfnamefont {M.}~\bibnamefont {Decker}},
  \bibinfo {author} {\bibfnamefont {E.}~\bibnamefont {Rusak}}, \bibinfo
  {author} {\bibfnamefont {D.~N.}\ \bibnamefont {Neshev}}, \ and\ \bibinfo
  {author} {\bibfnamefont {I.}~\bibnamefont {Brener}},\ }\href@noop {}
  {\bibfield  {journal} {\bibinfo  {journal} {ACS Nano}\ ,\ \bibinfo {pages}
  {4308}} (\bibinfo {year} {2015})}\BibitemShut {NoStop}%
\bibitem [{\citenamefont {Kats}\ \emph {et~al.}(2012)\citenamefont {Kats},
  \citenamefont {Sharma}, \citenamefont {Lin}, \citenamefont {Genevet},
  \citenamefont {Blanchard}, \citenamefont {Yang}, \citenamefont {Qazilbash},
  \citenamefont {Basov}, \citenamefont {Ramanathan},\ and\ \citenamefont
  {Capasso}}]{Kats2012}%
  \BibitemOpen
  \bibfield  {author} {\bibinfo {author} {\bibfnamefont {M.~A.}\ \bibnamefont
  {Kats}}, \bibinfo {author} {\bibfnamefont {D.}~\bibnamefont {Sharma}},
  \bibinfo {author} {\bibfnamefont {J.}~\bibnamefont {Lin}}, \bibinfo {author}
  {\bibfnamefont {P.}~\bibnamefont {Genevet}}, \bibinfo {author} {\bibfnamefont
  {R.}~\bibnamefont {Blanchard}}, \bibinfo {author} {\bibfnamefont
  {Z.}~\bibnamefont {Yang}}, \bibinfo {author} {\bibfnamefont {M.~M.}\
  \bibnamefont {Qazilbash}}, \bibinfo {author} {\bibfnamefont {D.~N.}\
  \bibnamefont {Basov}}, \bibinfo {author} {\bibfnamefont {S.}~\bibnamefont
  {Ramanathan}}, \ and\ \bibinfo {author} {\bibfnamefont {F.}~\bibnamefont
  {Capasso}},\ }\href {\doibase 10.1063/1.4767646} {\bibfield  {journal}
  {\bibinfo  {journal} {Applied Physics Letters}\ }\textbf {\bibinfo {volume}
  {101}} (\bibinfo {year} {2012}),\ 10.1063/1.4767646}\BibitemShut {NoStop}%
\bibitem [{\citenamefont {Rensberg}\ \emph {et~al.}(2016)\citenamefont
  {Rensberg}, \citenamefont {Zhang}, \citenamefont {Zhou}, \citenamefont
  {McLeod}, \citenamefont {Schwarz}, \citenamefont {Goldflam}, \citenamefont
  {Liu}, \citenamefont {Kerbusch}, \citenamefont {Nawrodt}, \citenamefont
  {Ramanathan}, \citenamefont {Basov}, \citenamefont {Capasso}, \citenamefont
  {Ronning},\ and\ \citenamefont {Kats}}]{Rensberg2016}%
  \BibitemOpen
  \bibfield  {author} {\bibinfo {author} {\bibfnamefont {J.}~\bibnamefont
  {Rensberg}}, \bibinfo {author} {\bibfnamefont {S.}~\bibnamefont {Zhang}},
  \bibinfo {author} {\bibfnamefont {Y.}~\bibnamefont {Zhou}}, \bibinfo {author}
  {\bibfnamefont {A.~S.}\ \bibnamefont {McLeod}}, \bibinfo {author}
  {\bibfnamefont {C.}~\bibnamefont {Schwarz}}, \bibinfo {author} {\bibfnamefont
  {M.}~\bibnamefont {Goldflam}}, \bibinfo {author} {\bibfnamefont
  {M.}~\bibnamefont {Liu}}, \bibinfo {author} {\bibfnamefont {J.}~\bibnamefont
  {Kerbusch}}, \bibinfo {author} {\bibfnamefont {R.}~\bibnamefont {Nawrodt}},
  \bibinfo {author} {\bibfnamefont {S.}~\bibnamefont {Ramanathan}}, \bibinfo
  {author} {\bibfnamefont {D.~N.}\ \bibnamefont {Basov}}, \bibinfo {author}
  {\bibfnamefont {F.}~\bibnamefont {Capasso}}, \bibinfo {author} {\bibfnamefont
  {C.}~\bibnamefont {Ronning}}, \ and\ \bibinfo {author} {\bibfnamefont
  {M.~A.}\ \bibnamefont {Kats}},\ }\href {\doibase
  10.1021/acs.nanolett.5b04122} {\bibfield  {journal} {\bibinfo  {journal}
  {Nano Letters}\ }\textbf {\bibinfo {volume} {16}},\ \bibinfo {pages} {1050}
  (\bibinfo {year} {2016})}\BibitemShut {NoStop}%
\bibitem [{\citenamefont {Wang}\ \emph {et~al.}(2015)\citenamefont {Wang},
  \citenamefont {Zhang}, \citenamefont {Gu}, \citenamefont {Mehmood},
  \citenamefont {Gong}, \citenamefont {Srivastava}, \citenamefont {Jian},
  \citenamefont {Venkatesan}, \citenamefont {Qiu},\ and\ \citenamefont
  {Hong}}]{Wang2015}%
  \BibitemOpen
  \bibfield  {author} {\bibinfo {author} {\bibfnamefont {D.}~\bibnamefont
  {Wang}}, \bibinfo {author} {\bibfnamefont {L.}~\bibnamefont {Zhang}},
  \bibinfo {author} {\bibfnamefont {Y.}~\bibnamefont {Gu}}, \bibinfo {author}
  {\bibfnamefont {M.~Q.}\ \bibnamefont {Mehmood}}, \bibinfo {author}
  {\bibfnamefont {Y.}~\bibnamefont {Gong}}, \bibinfo {author} {\bibfnamefont
  {A.}~\bibnamefont {Srivastava}}, \bibinfo {author} {\bibfnamefont
  {L.}~\bibnamefont {Jian}}, \bibinfo {author} {\bibfnamefont {T.}~\bibnamefont
  {Venkatesan}}, \bibinfo {author} {\bibfnamefont {C.~W.}\ \bibnamefont {Qiu}},
  \ and\ \bibinfo {author} {\bibfnamefont {M.}~\bibnamefont {Hong}},\ }\href
  {\doibase 10.1038/srep15020} {\bibfield  {journal} {\bibinfo  {journal} {Sci
  Rep}\ }\textbf {\bibinfo {volume} {5}},\ \bibinfo {pages} {15020} (\bibinfo
  {year} {2015})}\BibitemShut {NoStop}%
\bibitem [{\citenamefont {Waters}\ \emph {et~al.}(2015)\citenamefont {Waters},
  \citenamefont {Hobson}, \citenamefont {MacDonald},\ and\ \citenamefont
  {Zheludev}}]{Waters2015}%
  \BibitemOpen
  \bibfield  {author} {\bibinfo {author} {\bibfnamefont {R.~F.}\ \bibnamefont
  {Waters}}, \bibinfo {author} {\bibfnamefont {P.~A.}\ \bibnamefont {Hobson}},
  \bibinfo {author} {\bibfnamefont {K.~F.}\ \bibnamefont {MacDonald}}, \ and\
  \bibinfo {author} {\bibfnamefont {N.~I.}\ \bibnamefont {Zheludev}},\ }\href
  {\doibase 10.1063/1.4929396} {\bibfield  {journal} {\bibinfo  {journal}
  {Applied Physics Letters}\ }\textbf {\bibinfo {volume} {107}} (\bibinfo
  {year} {2015}),\ 10.1063/1.4929396}\BibitemShut {NoStop}%
\bibitem [{\citenamefont {Capasso}(2013)}]{Capasso2013}%
  \BibitemOpen
  \bibfield  {author} {\bibinfo {author} {\bibfnamefont {F.}~\bibnamefont
  {Capasso}},\ }\href
  {http://pubs.acs.org/doi/abs/10.1021/nl3047943{\%}5CnD:{\%}5CAppData{\%}5CRoaming{\%}5CMozilla{\%}5CFirefox{\%}5CProfiles{\%}5Cvw6qxvil.default{\%}5Czotero{\%}5Cstorage{\%}5CG2S6GVXE{\%}5Cnl3047943.html}
  {\bibfield  {journal} {\bibinfo  {journal} {Nano Letters}\ }\textbf {\bibinfo
  {volume} {13}},\ \bibinfo {pages} {1257} (\bibinfo {year}
  {2013})}\BibitemShut {NoStop}%
\bibitem [{\citenamefont {Fallahi}\ and\ \citenamefont
  {Perruisseau-Carrier}(2012)}]{Fallahi2012}%
  \BibitemOpen
  \bibfield  {author} {\bibinfo {author} {\bibfnamefont {A.}~\bibnamefont
  {Fallahi}}\ and\ \bibinfo {author} {\bibfnamefont {J.}~\bibnamefont
  {Perruisseau-Carrier}},\ }\href {\doibase 10.1103/PhysRevB.86.195408}
  {\bibfield  {journal} {\bibinfo  {journal} {Physical Review B - Condensed
  Matter and Materials Physics}\ }\textbf {\bibinfo {volume} {86}},\ \bibinfo
  {pages} {1} (\bibinfo {year} {2012})},\ \Eprint
  {http://arxiv.org/abs/1210.5611} {arXiv:1210.5611} \BibitemShut {NoStop}%
\bibitem [{\citenamefont {Yao}\ \emph {et~al.}(2014)\citenamefont {Yao},
  \citenamefont {Shankar}, \citenamefont {Kats}, \citenamefont {Song},
  \citenamefont {Kong}, \citenamefont {Loncar},\ and\ \citenamefont
  {Capasso}}]{Yao2014}%
  \BibitemOpen
  \bibfield  {author} {\bibinfo {author} {\bibfnamefont {Y.}~\bibnamefont
  {Yao}}, \bibinfo {author} {\bibfnamefont {R.}~\bibnamefont {Shankar}},
  \bibinfo {author} {\bibfnamefont {M.~A.}\ \bibnamefont {Kats}}, \bibinfo
  {author} {\bibfnamefont {Y.}~\bibnamefont {Song}}, \bibinfo {author}
  {\bibfnamefont {J.}~\bibnamefont {Kong}}, \bibinfo {author} {\bibfnamefont
  {M.}~\bibnamefont {Loncar}}, \ and\ \bibinfo {author} {\bibfnamefont
  {F.}~\bibnamefont {Capasso}},\ }\href {\doibase 10.1021/nl503104n} {\bibfield
   {journal} {\bibinfo  {journal} {Nano Letters}\ }\textbf {\bibinfo {volume}
  {14}},\ \bibinfo {pages} {6526} (\bibinfo {year} {2014})}\BibitemShut
  {NoStop}%
\bibitem [{\citenamefont {{Z. Fang}}\ \emph {et~al.}(2013)\citenamefont {{Z.
  Fang}}, \citenamefont {{S. Thongrattanasiri}}, \citenamefont {{A.
  Schlather}}, \citenamefont {{Z. Liu}}, \citenamefont {{L. Ma}}, \citenamefont
  {{Y. Wang}}, \citenamefont {{P. M. Ajayan}}, \citenamefont {{P. Nordlander}},
  \citenamefont {{N. J. Halas}},\ and\ \citenamefont {{F. J. Garcia de
  Abajo}}}]{Z.Fang2013}%
  \BibitemOpen
  \bibfield  {author} {\bibinfo {author} {\bibnamefont {{Z. Fang}}}, \bibinfo
  {author} {\bibnamefont {{S. Thongrattanasiri}}}, \bibinfo {author}
  {\bibnamefont {{A. Schlather}}}, \bibinfo {author} {\bibnamefont {{Z. Liu}}},
  \bibinfo {author} {\bibnamefont {{L. Ma}}}, \bibinfo {author} {\bibnamefont
  {{Y. Wang}}}, \bibinfo {author} {\bibnamefont {{P. M. Ajayan}}}, \bibinfo
  {author} {\bibnamefont {{P. Nordlander}}}, \bibinfo {author} {\bibnamefont
  {{N. J. Halas}}}, \ and\ \bibinfo {author} {\bibnamefont {{F. J. Garcia de
  Abajo}}},\ }\href {\doibase 10.1021/nn3055835} {\bibfield  {journal}
  {\bibinfo  {journal} {ACS Nano}\ }\textbf {\bibinfo {volume} {7}},\ \bibinfo
  {pages} {2388�2395} (\bibinfo {year} {2013})}\BibitemShut {NoStop}%
\bibitem [{\citenamefont {Leroux}\ \emph {et~al.}(2009)\citenamefont {Leroux},
  \citenamefont {Lacroix}, \citenamefont {Fave}, \citenamefont {Stockhausen},
  \citenamefont {Felidj}, \citenamefont {Grand}, \citenamefont {Hohenau},\ and\
  \citenamefont {Krenn}}]{Leroux2009}%
  \BibitemOpen
  \bibfield  {author} {\bibinfo {author} {\bibfnamefont {Y.}~\bibnamefont
  {Leroux}}, \bibinfo {author} {\bibfnamefont {J.~C.}\ \bibnamefont {Lacroix}},
  \bibinfo {author} {\bibfnamefont {C.}~\bibnamefont {Fave}}, \bibinfo {author}
  {\bibfnamefont {V.}~\bibnamefont {Stockhausen}}, \bibinfo {author}
  {\bibfnamefont {N.}~\bibnamefont {Felidj}}, \bibinfo {author} {\bibfnamefont
  {J.}~\bibnamefont {Grand}}, \bibinfo {author} {\bibfnamefont
  {A.}~\bibnamefont {Hohenau}}, \ and\ \bibinfo {author} {\bibfnamefont
  {J.~R.}\ \bibnamefont {Krenn}},\ }\href {\doibase 10.1021/nl900695j}
  {\bibfield  {journal} {\bibinfo  {journal} {Nano Letters}\ }\textbf {\bibinfo
  {volume} {9}},\ \bibinfo {pages} {2144} (\bibinfo {year} {2009})}\BibitemShut
  {NoStop}%
\bibitem [{\citenamefont {Sun}\ \emph {et~al.}(2013)\citenamefont {Sun},
  \citenamefont {Timurdogan}, \citenamefont {Yaacobi}, \citenamefont
  {Hosseini},\ and\ \citenamefont {Watts}}]{Sun2013}%
  \BibitemOpen
  \bibfield  {author} {\bibinfo {author} {\bibfnamefont {J.}~\bibnamefont
  {Sun}}, \bibinfo {author} {\bibfnamefont {E.}~\bibnamefont {Timurdogan}},
  \bibinfo {author} {\bibfnamefont {A.}~\bibnamefont {Yaacobi}}, \bibinfo
  {author} {\bibfnamefont {E.~S.}\ \bibnamefont {Hosseini}}, \ and\ \bibinfo
  {author} {\bibfnamefont {M.~R.}\ \bibnamefont {Watts}},\ }\href {\doibase
  10.1038/nature11727} {\bibfield  {journal} {\bibinfo  {journal} {Nature}\
  }\textbf {\bibinfo {volume} {493}},\ \bibinfo {pages} {195} (\bibinfo {year}
  {2013})}\BibitemShut {NoStop}%
\bibitem [{\citenamefont {Pryce}\ \emph {et~al.}(2010)\citenamefont {Pryce},
  \citenamefont {Aydin}, \citenamefont {Kelaita}, \citenamefont {Briggs},\ and\
  \citenamefont {Atwater}}]{Pryce2010}%
  \BibitemOpen
  \bibfield  {author} {\bibinfo {author} {\bibfnamefont {I.~M.}\ \bibnamefont
  {Pryce}}, \bibinfo {author} {\bibfnamefont {K.}~\bibnamefont {Aydin}},
  \bibinfo {author} {\bibfnamefont {Y.~A.}\ \bibnamefont {Kelaita}}, \bibinfo
  {author} {\bibfnamefont {R.~M.}\ \bibnamefont {Briggs}}, \ and\ \bibinfo
  {author} {\bibfnamefont {H.~A.}\ \bibnamefont {Atwater}},\ }\href {\doibase
  10.1021/nl102684x} {\bibfield  {journal} {\bibinfo  {journal} {Nano Letters}\
  }\textbf {\bibinfo {volume} {10}},\ \bibinfo {pages} {4222} (\bibinfo {year}
  {2010})}\BibitemShut {NoStop}%
\bibitem [{\citenamefont {Zheludev}\ and\ \citenamefont
  {Plum}(2016)}]{Zheludev2016}%
  \BibitemOpen
  \bibfield  {author} {\bibinfo {author} {\bibfnamefont {N.~I.}\ \bibnamefont
  {Zheludev}}\ and\ \bibinfo {author} {\bibfnamefont {E.}~\bibnamefont
  {Plum}},\ }\href {\doibase
  10.1038/nnano.2015.302\rhttp://www.nature.com/nnano/journal/v11/n1/abs/nnano.2015.302.html#supplementary-information}
  {\bibfield  {journal} {\bibinfo  {journal} {Nat Nano}\ }\textbf {\bibinfo
  {volume} {11}},\ \bibinfo {pages} {16} (\bibinfo {year} {2016})}\BibitemShut
  {NoStop}%
\bibitem [{\citenamefont {Chen}\ \emph {et~al.}(2017)\citenamefont {Chen},
  \citenamefont {{\"O}zdemir}, \citenamefont {Zhao}, \citenamefont {Wiersig},\
  and\ \citenamefont {Yang}}]{LanNature2017}%
  \BibitemOpen
  \bibfield  {author} {\bibinfo {author} {\bibfnamefont {W.}~\bibnamefont
  {Chen}}, \bibinfo {author} {\bibfnamefont {{\c{S}}.~K.}\ \bibnamefont
  {{\"O}zdemir}}, \bibinfo {author} {\bibfnamefont {G.}~\bibnamefont {Zhao}},
  \bibinfo {author} {\bibfnamefont {J.}~\bibnamefont {Wiersig}}, \ and\
  \bibinfo {author} {\bibfnamefont {L.}~\bibnamefont {Yang}},\ }\href@noop {}
  {\bibfield  {journal} {\bibinfo  {journal} {Nature}\ }\textbf {\bibinfo
  {volume} {548}},\ \bibinfo {pages} {192} (\bibinfo {year}
  {2017})}\BibitemShut {NoStop}%
\bibitem [{\citenamefont {Feng}\ \emph {et~al.}(2013)\citenamefont {Feng},
  \citenamefont {Xu}, \citenamefont {Fegadolli}, \citenamefont {Lu},
  \citenamefont {Oliveira}, \citenamefont {Almeida}, \citenamefont {Chen},\
  and\ \citenamefont {Scherer}}]{Feng:2012jj}%
  \BibitemOpen
  \bibfield  {author} {\bibinfo {author} {\bibfnamefont {L.}~\bibnamefont
  {Feng}}, \bibinfo {author} {\bibfnamefont {Y.-L.}\ \bibnamefont {Xu}},
  \bibinfo {author} {\bibfnamefont {W.~S.}\ \bibnamefont {Fegadolli}}, \bibinfo
  {author} {\bibfnamefont {M.-H.}\ \bibnamefont {Lu}}, \bibinfo {author}
  {\bibfnamefont {J.~E.~B.}\ \bibnamefont {Oliveira}}, \bibinfo {author}
  {\bibfnamefont {V.~R.}\ \bibnamefont {Almeida}}, \bibinfo {author}
  {\bibfnamefont {Y.-F.}\ \bibnamefont {Chen}}, \ and\ \bibinfo {author}
  {\bibfnamefont {A.}~\bibnamefont {Scherer}},\ }\href@noop {} {\bibfield
  {journal} {\bibinfo  {journal} {Nature Materials}\ }\textbf {\bibinfo
  {volume} {12}},\ \bibinfo {pages} {108} (\bibinfo {year} {2013})}\BibitemShut
  {NoStop}%
\bibitem [{\citenamefont {Feng}\ \emph {et~al.}(2014)\citenamefont {Feng},
  \citenamefont {Wong}, \citenamefont {Ma}, \citenamefont {Wang},\ and\
  \citenamefont {Zhang}}]{Feng:2014gg}%
  \BibitemOpen
  \bibfield  {author} {\bibinfo {author} {\bibfnamefont {L.}~\bibnamefont
  {Feng}}, \bibinfo {author} {\bibfnamefont {Z.~J.}\ \bibnamefont {Wong}},
  \bibinfo {author} {\bibfnamefont {R.-M.}\ \bibnamefont {Ma}}, \bibinfo
  {author} {\bibfnamefont {Y.}~\bibnamefont {Wang}}, \ and\ \bibinfo {author}
  {\bibfnamefont {X.}~\bibnamefont {Zhang}},\ }\href@noop {} {\bibfield
  {journal} {\bibinfo  {journal} {Science}\ }\textbf {\bibinfo {volume}
  {346}},\ \bibinfo {pages} {972} (\bibinfo {year} {2014})}\BibitemShut
  {NoStop}%
\bibitem [{\citenamefont {Baum}\ \emph {et~al.}(2015)\citenamefont {Baum},
  \citenamefont {Alaeian},\ and\ \citenamefont {Dionne}}]{Baum:2015id}%
  \BibitemOpen
  \bibfield  {author} {\bibinfo {author} {\bibfnamefont {B.}~\bibnamefont
  {Baum}}, \bibinfo {author} {\bibfnamefont {H.}~\bibnamefont {Alaeian}}, \
  and\ \bibinfo {author} {\bibfnamefont {J.}~\bibnamefont {Dionne}},\
  }\href@noop {} {\bibfield  {journal} {\bibinfo  {journal} {Journal of Applied
  Physics}\ }\textbf {\bibinfo {volume} {117}},\ \bibinfo {pages} {063106}
  (\bibinfo {year} {2015})}\BibitemShut {NoStop}%
\bibitem [{\citenamefont {Ramezani}\ \emph {et~al.}(2010)\citenamefont
  {Ramezani}, \citenamefont {Kottos}, \citenamefont {El-Ganainy},\ and\
  \citenamefont {Christodoulides}}]{Ramezani:2010eb}%
  \BibitemOpen
  \bibfield  {author} {\bibinfo {author} {\bibfnamefont {H.}~\bibnamefont
  {Ramezani}}, \bibinfo {author} {\bibfnamefont {T.}~\bibnamefont {Kottos}},
  \bibinfo {author} {\bibfnamefont {R.}~\bibnamefont {El-Ganainy}}, \ and\
  \bibinfo {author} {\bibfnamefont {D.~N.}\ \bibnamefont {Christodoulides}},\
  }\href@noop {} {\bibfield  {journal} {\bibinfo  {journal} {Physical Review
  A}\ }\textbf {\bibinfo {volume} {82}},\ \bibinfo {pages} {043803} (\bibinfo
  {year} {2010})}\BibitemShut {NoStop}%
\bibitem [{\citenamefont {Peng}\ \emph {et~al.}(2014)\citenamefont {Peng},
  \citenamefont {Ozdemir}, \citenamefont {Lei}, \citenamefont {Monifi},
  \citenamefont {Gianfreda}, \citenamefont {Long}, \citenamefont {Fan},
  \citenamefont {Nori}, \citenamefont {Bender},\ and\ \citenamefont
  {Yang}}]{Peng:2014kl}%
  \BibitemOpen
  \bibfield  {author} {\bibinfo {author} {\bibfnamefont {B.}~\bibnamefont
  {Peng}}, \bibinfo {author} {\bibfnamefont {S.~K.}\ \bibnamefont {Ozdemir}},
  \bibinfo {author} {\bibfnamefont {F.}~\bibnamefont {Lei}}, \bibinfo {author}
  {\bibfnamefont {F.}~\bibnamefont {Monifi}}, \bibinfo {author} {\bibfnamefont
  {M.}~\bibnamefont {Gianfreda}}, \bibinfo {author} {\bibfnamefont {G.~L.}\
  \bibnamefont {Long}}, \bibinfo {author} {\bibfnamefont {S.}~\bibnamefont
  {Fan}}, \bibinfo {author} {\bibfnamefont {F.}~\bibnamefont {Nori}}, \bibinfo
  {author} {\bibfnamefont {C.~M.}\ \bibnamefont {Bender}}, \ and\ \bibinfo
  {author} {\bibfnamefont {L.}~\bibnamefont {Yang}},\ }\href@noop {} {\bibfield
   {journal} {\bibinfo  {journal} {Nature Physics}\ } (\bibinfo {year}
  {2014})}\BibitemShut {NoStop}%
\bibitem [{\citenamefont {Lawrence}\ \emph {et~al.}(2014)\citenamefont
  {Lawrence}, \citenamefont {Xu}, \citenamefont {Zhang}, \citenamefont {Cong},
  \citenamefont {Han}, \citenamefont {Zhang},\ and\ \citenamefont
  {Zhang}}]{Lawrence2014}%
  \BibitemOpen
  \bibfield  {author} {\bibinfo {author} {\bibfnamefont {M.}~\bibnamefont
  {Lawrence}}, \bibinfo {author} {\bibfnamefont {N.}~\bibnamefont {Xu}},
  \bibinfo {author} {\bibfnamefont {X.}~\bibnamefont {Zhang}}, \bibinfo
  {author} {\bibfnamefont {L.}~\bibnamefont {Cong}}, \bibinfo {author}
  {\bibfnamefont {J.}~\bibnamefont {Han}}, \bibinfo {author} {\bibfnamefont
  {W.}~\bibnamefont {Zhang}}, \ and\ \bibinfo {author} {\bibfnamefont
  {S.}~\bibnamefont {Zhang}},\ }\href {\doibase 10.1103/PhysRevLett.113.093901}
  {\bibfield  {journal} {\bibinfo  {journal} {Physical Review Letters}\
  }\textbf {\bibinfo {volume} {113}},\ \bibinfo {pages} {1} (\bibinfo {year}
  {2014})}\BibitemShut {NoStop}%
\bibitem [{\citenamefont {Alaeian}\ and\ \citenamefont
  {Dionne}(2014{\natexlab{a}})}]{Alaeian:2014dj}%
  \BibitemOpen
  \bibfield  {author} {\bibinfo {author} {\bibfnamefont {H.}~\bibnamefont
  {Alaeian}}\ and\ \bibinfo {author} {\bibfnamefont {J.~A.}\ \bibnamefont
  {Dionne}},\ }\href@noop {} {\bibfield  {journal} {\bibinfo  {journal}
  {Physical Review A}\ }\textbf {\bibinfo {volume} {89}},\ \bibinfo {pages}
  {33829} (\bibinfo {year} {2014}{\natexlab{a}})}\BibitemShut {NoStop}%
\bibitem [{\citenamefont {Benisty}\ \emph {et~al.}(2011)\citenamefont
  {Benisty}, \citenamefont {Degiron}, \citenamefont {Lupu}, \citenamefont
  {De~Lustrac}, \citenamefont {Ch{\'e}nais}, \citenamefont {Forget},
  \citenamefont {Besbes}, \citenamefont {Barbillon}, \citenamefont {Bruyant},\
  and\ \citenamefont {Blaize}}]{Benisty:2011wq}%
  \BibitemOpen
  \bibfield  {author} {\bibinfo {author} {\bibfnamefont {H.}~\bibnamefont
  {Benisty}}, \bibinfo {author} {\bibfnamefont {A.}~\bibnamefont {Degiron}},
  \bibinfo {author} {\bibfnamefont {A.}~\bibnamefont {Lupu}}, \bibinfo {author}
  {\bibfnamefont {A.}~\bibnamefont {De~Lustrac}}, \bibinfo {author}
  {\bibfnamefont {S.}~\bibnamefont {Ch{\'e}nais}}, \bibinfo {author}
  {\bibfnamefont {S.}~\bibnamefont {Forget}}, \bibinfo {author} {\bibfnamefont
  {M.}~\bibnamefont {Besbes}}, \bibinfo {author} {\bibfnamefont
  {G.}~\bibnamefont {Barbillon}}, \bibinfo {author} {\bibfnamefont
  {A.}~\bibnamefont {Bruyant}}, \ and\ \bibinfo {author} {\bibfnamefont
  {S.}~\bibnamefont {Blaize}},\ }\href@noop {} {\bibfield  {journal} {\bibinfo
  {journal} {Optics Express}\ }\textbf {\bibinfo {volume} {19}},\ \bibinfo
  {pages} {18004} (\bibinfo {year} {2011})}\BibitemShut {NoStop}%
\bibitem [{\citenamefont {Alaeian}\ and\ \citenamefont
  {Dionne}(2014{\natexlab{b}})}]{Alaeian:2014eb}%
  \BibitemOpen
  \bibfield  {author} {\bibinfo {author} {\bibfnamefont {H.}~\bibnamefont
  {Alaeian}}\ and\ \bibinfo {author} {\bibfnamefont {J.~A.}\ \bibnamefont
  {Dionne}},\ }\href@noop {} {\bibfield  {journal} {\bibinfo  {journal}
  {Physical Review B}\ }\textbf {\bibinfo {volume} {89}},\ \bibinfo {pages}
  {75136} (\bibinfo {year} {2014}{\natexlab{b}})}\BibitemShut {NoStop}%
\bibitem [{\citenamefont {Guo}\ \emph {et~al.}(2009)\citenamefont {Guo},
  \citenamefont {Salamo}, \citenamefont {Duchesne}, \citenamefont {Morandotti},
  \citenamefont {Volatier-Ravat}, \citenamefont {Aimez}, \citenamefont
  {Siviloglou},\ and\ \citenamefont {Christodoulides}}]{Guo:2009hd}%
  \BibitemOpen
  \bibfield  {author} {\bibinfo {author} {\bibfnamefont {A.}~\bibnamefont
  {Guo}}, \bibinfo {author} {\bibfnamefont {G.~J.}\ \bibnamefont {Salamo}},
  \bibinfo {author} {\bibfnamefont {D.}~\bibnamefont {Duchesne}}, \bibinfo
  {author} {\bibfnamefont {R.}~\bibnamefont {Morandotti}}, \bibinfo {author}
  {\bibfnamefont {M.}~\bibnamefont {Volatier-Ravat}}, \bibinfo {author}
  {\bibfnamefont {V.}~\bibnamefont {Aimez}}, \bibinfo {author} {\bibfnamefont
  {G.~A.}\ \bibnamefont {Siviloglou}}, \ and\ \bibinfo {author} {\bibfnamefont
  {D.~N.}\ \bibnamefont {Christodoulides}},\ }\href@noop {} {\bibfield
  {journal} {\bibinfo  {journal} {Physical Review Letters}\ }\textbf {\bibinfo
  {volume} {103}},\ \bibinfo {pages} {093902} (\bibinfo {year}
  {2009})}\BibitemShut {NoStop}%
\bibitem [{\citenamefont {Yu}\ \emph {et~al.}(2016)\citenamefont {Yu},
  \citenamefont {Piao},\ and\ \citenamefont {Park}}]{yu2016acceleration}%
  \BibitemOpen
  \bibfield  {author} {\bibinfo {author} {\bibfnamefont {S.}~\bibnamefont
  {Yu}}, \bibinfo {author} {\bibfnamefont {X.}~\bibnamefont {Piao}}, \ and\
  \bibinfo {author} {\bibfnamefont {N.}~\bibnamefont {Park}},\ }\href@noop {}
  {\bibfield  {journal} {\bibinfo  {journal} {Scientific reports}\ }\textbf
  {\bibinfo {volume} {6}},\ \bibinfo {pages} {37754} (\bibinfo {year}
  {2016})}\BibitemShut {NoStop}%
\bibitem [{\citenamefont {Cerjan}\ and\ \citenamefont
  {Fan}(2017)}]{cerjan2017achieving}%
  \BibitemOpen
  \bibfield  {author} {\bibinfo {author} {\bibfnamefont {A.}~\bibnamefont
  {Cerjan}}\ and\ \bibinfo {author} {\bibfnamefont {S.}~\bibnamefont {Fan}},\
  }\href@noop {} {\bibfield  {journal} {\bibinfo  {journal} {Physical review
  letters}\ }\textbf {\bibinfo {volume} {118}},\ \bibinfo {pages} {253902}
  (\bibinfo {year} {2017})}\BibitemShut {NoStop}%
\bibitem [{\citenamefont {Hassan}\ \emph {et~al.}(2017)\citenamefont {Hassan},
  \citenamefont {Zhen}, \citenamefont {Solja{\v{c}}i{\'c}}, \citenamefont
  {Khajavikhan},\ and\ \citenamefont
  {Christodoulides}}]{hassan2017dynamically}%
  \BibitemOpen
  \bibfield  {author} {\bibinfo {author} {\bibfnamefont {A.~U.}\ \bibnamefont
  {Hassan}}, \bibinfo {author} {\bibfnamefont {B.}~\bibnamefont {Zhen}},
  \bibinfo {author} {\bibfnamefont {M.}~\bibnamefont {Solja{\v{c}}i{\'c}}},
  \bibinfo {author} {\bibfnamefont {M.}~\bibnamefont {Khajavikhan}}, \ and\
  \bibinfo {author} {\bibfnamefont {D.~N.}\ \bibnamefont {Christodoulides}},\
  }\href@noop {} {\bibfield  {journal} {\bibinfo  {journal} {Physical review
  letters}\ }\textbf {\bibinfo {volume} {118}},\ \bibinfo {pages} {093002}
  (\bibinfo {year} {2017})}\BibitemShut {NoStop}%
\bibitem [{\citenamefont {Ge}\ and\ \citenamefont {Stone}(2014)}]{Ge:2014fd}%
  \BibitemOpen
  \bibfield  {author} {\bibinfo {author} {\bibfnamefont {L.}~\bibnamefont
  {Ge}}\ and\ \bibinfo {author} {\bibfnamefont {A.~D.}\ \bibnamefont {Stone}},\
  }\href@noop {} {\bibfield  {journal} {\bibinfo  {journal} {Physical Review
  X}\ }\textbf {\bibinfo {volume} {4}},\ \bibinfo {pages} {31011} (\bibinfo
  {year} {2014})}\BibitemShut {NoStop}%
\bibitem [{\citenamefont {Alaeian}\ \emph {et~al.}(2016)\citenamefont
  {Alaeian}, \citenamefont {Baum}, \citenamefont {Jankovic}, \citenamefont
  {Lawrence},\ and\ \citenamefont {Dionne}}]{Alaeian:2016hr}%
  \BibitemOpen
  \bibfield  {author} {\bibinfo {author} {\bibfnamefont {H.}~\bibnamefont
  {Alaeian}}, \bibinfo {author} {\bibfnamefont {B.}~\bibnamefont {Baum}},
  \bibinfo {author} {\bibfnamefont {V.}~\bibnamefont {Jankovic}}, \bibinfo
  {author} {\bibfnamefont {M.}~\bibnamefont {Lawrence}}, \ and\ \bibinfo
  {author} {\bibfnamefont {J.~A.}\ \bibnamefont {Dionne}},\ }\href@noop {}
  {\bibfield  {journal} {\bibinfo  {journal} {Physical Review B}\ }\textbf
  {\bibinfo {volume} {93}},\ \bibinfo {pages} {205439} (\bibinfo {year}
  {2016})}\BibitemShut {NoStop}%
\bibitem [{\citenamefont {Chitsazi}\ \emph {et~al.}(2017)\citenamefont
  {Chitsazi}, \citenamefont {Li}, \citenamefont {Ellis},\ and\ \citenamefont
  {Kottos}}]{Tsampikos2017}%
  \BibitemOpen
  \bibfield  {author} {\bibinfo {author} {\bibfnamefont {M.}~\bibnamefont
  {Chitsazi}}, \bibinfo {author} {\bibfnamefont {H.}~\bibnamefont {Li}},
  \bibinfo {author} {\bibfnamefont {F.}~\bibnamefont {Ellis}}, \ and\ \bibinfo
  {author} {\bibfnamefont {T.}~\bibnamefont {Kottos}},\ }\href@noop {}
  {\bibfield  {journal} {\bibinfo  {journal} {Physical review letters}\
  }\textbf {\bibinfo {volume} {119}},\ \bibinfo {pages} {093901} (\bibinfo
  {year} {2017})}\BibitemShut {NoStop}%
\bibitem [{\citenamefont {Kodigala}\ \emph {et~al.}(2016)\citenamefont
  {Kodigala}, \citenamefont {Lepetit},\ and\ \citenamefont
  {Kant{\'e}}}]{Kante16}%
  \BibitemOpen
  \bibfield  {author} {\bibinfo {author} {\bibfnamefont {A.}~\bibnamefont
  {Kodigala}}, \bibinfo {author} {\bibfnamefont {T.}~\bibnamefont {Lepetit}}, \
  and\ \bibinfo {author} {\bibfnamefont {B.}~\bibnamefont {Kant{\'e}}},\
  }\href@noop {} {\bibfield  {journal} {\bibinfo  {journal} {Physical Review
  B}\ }\textbf {\bibinfo {volume} {94}},\ \bibinfo {pages} {201103} (\bibinfo
  {year} {2016})}\BibitemShut {NoStop}%
\bibitem [{\citenamefont {Baida}\ and\ \citenamefont
  {Van~Labeke}(2003)}]{Baida:2003gy}%
  \BibitemOpen
  \bibfield  {author} {\bibinfo {author} {\bibfnamefont {F.}~\bibnamefont
  {Baida}}\ and\ \bibinfo {author} {\bibfnamefont {D.}~\bibnamefont
  {Van~Labeke}},\ }\href@noop {} {\bibfield  {journal} {\bibinfo  {journal}
  {Physical Review B}\ }\textbf {\bibinfo {volume} {67}},\ \bibinfo {pages}
  {155314} (\bibinfo {year} {2003})}\BibitemShut {NoStop}%
\bibitem [{\citenamefont {Baida}\ \emph {et~al.}(2006)\citenamefont {Baida},
  \citenamefont {Belkhir}, \citenamefont {Van~Labeke},\ and\ \citenamefont
  {Lamrous}}]{Baida:2006fi}%
  \BibitemOpen
  \bibfield  {author} {\bibinfo {author} {\bibfnamefont {F.}~\bibnamefont
  {Baida}}, \bibinfo {author} {\bibfnamefont {A.}~\bibnamefont {Belkhir}},
  \bibinfo {author} {\bibfnamefont {D.}~\bibnamefont {Van~Labeke}}, \ and\
  \bibinfo {author} {\bibfnamefont {O.}~\bibnamefont {Lamrous}},\ }\href@noop
  {} {\bibfield  {journal} {\bibinfo  {journal} {Physical Review B}\ }\textbf
  {\bibinfo {volume} {74}},\ \bibinfo {pages} {205419} (\bibinfo {year}
  {2006})}\BibitemShut {NoStop}%
\bibitem [{\citenamefont {Poujet}\ \emph {et~al.}(2007)\citenamefont {Poujet},
  \citenamefont {Salvi},\ and\ \citenamefont {Baida}}]{Poujet:2007bg}%
  \BibitemOpen
  \bibfield  {author} {\bibinfo {author} {\bibfnamefont {Y.}~\bibnamefont
  {Poujet}}, \bibinfo {author} {\bibfnamefont {J.}~\bibnamefont {Salvi}}, \
  and\ \bibinfo {author} {\bibfnamefont {F.~I.}\ \bibnamefont {Baida}},\
  }\href@noop {} {\bibfield  {journal} {\bibinfo  {journal} {Optics Letters}\
  }\textbf {\bibinfo {volume} {32}},\ \bibinfo {pages} {2942} (\bibinfo {year}
  {2007})}\BibitemShut {NoStop}%
\bibitem [{\citenamefont {Orbons}\ \emph {et~al.}(2007)\citenamefont {Orbons},
  \citenamefont {Roberts}, \citenamefont {Jamieson}, \citenamefont {Haftel},
  \citenamefont {Schlockermann}, \citenamefont {Freeman},\ and\ \citenamefont
  {Luther-Davies}}]{Orbons:2007er}%
  \BibitemOpen
  \bibfield  {author} {\bibinfo {author} {\bibfnamefont {S.~M.}\ \bibnamefont
  {Orbons}}, \bibinfo {author} {\bibfnamefont {A.}~\bibnamefont {Roberts}},
  \bibinfo {author} {\bibfnamefont {D.~N.}\ \bibnamefont {Jamieson}}, \bibinfo
  {author} {\bibfnamefont {M.~I.}\ \bibnamefont {Haftel}}, \bibinfo {author}
  {\bibfnamefont {C.}~\bibnamefont {Schlockermann}}, \bibinfo {author}
  {\bibfnamefont {D.}~\bibnamefont {Freeman}}, \ and\ \bibinfo {author}
  {\bibfnamefont {B.}~\bibnamefont {Luther-Davies}},\ }\href@noop {} {\bibfield
   {journal} {\bibinfo  {journal} {Applied Physics Letters}\ }\textbf {\bibinfo
  {volume} {90}},\ \bibinfo {pages} {251107} (\bibinfo {year}
  {2007})}\BibitemShut {NoStop}%
\bibitem [{\citenamefont {Burgos}\ \emph {et~al.}(2010)\citenamefont {Burgos},
  \citenamefont {de~Waele}, \citenamefont {Polman},\ and\ \citenamefont
  {Atwater}}]{Burgos:2010ja}%
  \BibitemOpen
  \bibfield  {author} {\bibinfo {author} {\bibfnamefont {S.~P.}\ \bibnamefont
  {Burgos}}, \bibinfo {author} {\bibfnamefont {R.}~\bibnamefont {de~Waele}},
  \bibinfo {author} {\bibfnamefont {A.}~\bibnamefont {Polman}}, \ and\ \bibinfo
  {author} {\bibfnamefont {H.~A.}\ \bibnamefont {Atwater}},\ }\href@noop {}
  {\bibfield  {journal} {\bibinfo  {journal} {Nature Materials}\ }\textbf
  {\bibinfo {volume} {9}},\ \bibinfo {pages} {407} (\bibinfo {year}
  {2010})}\BibitemShut {NoStop}%
\bibitem [{\citenamefont {van~de Haar}\ \emph {et~al.}(2014)\citenamefont
  {van~de Haar}, \citenamefont {Maas}, \citenamefont {Schokker},\ and\
  \citenamefont {Polman}}]{vandeHaar:2014eu}%
  \BibitemOpen
  \bibfield  {author} {\bibinfo {author} {\bibfnamefont {M.~A.}\ \bibnamefont
  {van~de Haar}}, \bibinfo {author} {\bibfnamefont {R.}~\bibnamefont {Maas}},
  \bibinfo {author} {\bibfnamefont {H.}~\bibnamefont {Schokker}}, \ and\
  \bibinfo {author} {\bibfnamefont {A.}~\bibnamefont {Polman}},\ }\href@noop {}
  {\bibfield  {journal} {\bibinfo  {journal} {Nano Letters}\ }\textbf {\bibinfo
  {volume} {14}},\ \bibinfo {pages} {6356} (\bibinfo {year}
  {2014})}\BibitemShut {NoStop}%
\bibitem [{\citenamefont {Saleh}\ and\ \citenamefont
  {Dionne}(2012)}]{Saleh:2012em}%
  \BibitemOpen
  \bibfield  {author} {\bibinfo {author} {\bibfnamefont {A.~A.~E.}\
  \bibnamefont {Saleh}}\ and\ \bibinfo {author} {\bibfnamefont {J.~A.}\
  \bibnamefont {Dionne}},\ }\href@noop {} {\bibfield  {journal} {\bibinfo
  {journal} {Physical Review B}\ }\textbf {\bibinfo {volume} {85}},\ \bibinfo
  {pages} {045407} (\bibinfo {year} {2012})}\BibitemShut {NoStop}%
\bibitem [{\citenamefont {Khajavikhan}(2012)}]{Khajavikhan2012}%
  \BibitemOpen
  \bibfield  {author} {\bibinfo {author} {\bibfnamefont {M.}~\bibnamefont
  {Khajavikhan}},\ }\href@noop {} {\bibfield  {journal} {\bibinfo  {journal}
  {Nature}\ }\textbf {\bibinfo {volume} {482}},\ \bibinfo {pages} {204}
  (\bibinfo {year} {2012})}\BibitemShut {NoStop}%
\bibitem [{\citenamefont {Johnson}\ and\ \citenamefont
  {Christy}(1972)}]{PhysRevB.6.4370}%
  \BibitemOpen
  \bibfield  {author} {\bibinfo {author} {\bibfnamefont {P.~B.}\ \bibnamefont
  {Johnson}}\ and\ \bibinfo {author} {\bibfnamefont {R.~W.}\ \bibnamefont
  {Christy}},\ }\href@noop {} {\bibfield  {journal} {\bibinfo  {journal}
  {Physical Review B}\ }\textbf {\bibinfo {volume} {6}},\ \bibinfo {pages}
  {4370} (\bibinfo {year} {1972})}\BibitemShut {NoStop}%
\bibitem [{\citenamefont {Berini}\ and\ \citenamefont
  {De~Leon}(2012)}]{berini2012surface}%
  \BibitemOpen
  \bibfield  {author} {\bibinfo {author} {\bibfnamefont {P.}~\bibnamefont
  {Berini}}\ and\ \bibinfo {author} {\bibfnamefont {I.}~\bibnamefont
  {De~Leon}},\ }\href@noop {} {\bibfield  {journal} {\bibinfo  {journal}
  {Nature Photonics}\ }\textbf {\bibinfo {volume} {6}},\ \bibinfo {pages} {16}
  (\bibinfo {year} {2012})}\BibitemShut {NoStop}%
\bibitem [{\citenamefont {Xiao}\ \emph {et~al.}(2010)\citenamefont {Xiao},
  \citenamefont {Drachev}, \citenamefont {Kildishev}, \citenamefont {Ni},
  \citenamefont {Chettiar}, \citenamefont {Yuan},\ and\ \citenamefont
  {Shalaev}}]{xiao2010loss}%
  \BibitemOpen
  \bibfield  {author} {\bibinfo {author} {\bibfnamefont {S.}~\bibnamefont
  {Xiao}}, \bibinfo {author} {\bibfnamefont {V.~P.}\ \bibnamefont {Drachev}},
  \bibinfo {author} {\bibfnamefont {A.~V.}\ \bibnamefont {Kildishev}}, \bibinfo
  {author} {\bibfnamefont {X.}~\bibnamefont {Ni}}, \bibinfo {author}
  {\bibfnamefont {U.~K.}\ \bibnamefont {Chettiar}}, \bibinfo {author}
  {\bibfnamefont {H.-K.}\ \bibnamefont {Yuan}}, \ and\ \bibinfo {author}
  {\bibfnamefont {V.~M.}\ \bibnamefont {Shalaev}},\ }\href@noop {} {\bibfield
  {journal} {\bibinfo  {journal} {Nature}\ }\textbf {\bibinfo {volume} {466}},\
  \bibinfo {pages} {735} (\bibinfo {year} {2010})}\BibitemShut {NoStop}%
\bibitem [{\citenamefont {De~Leon}\ and\ \citenamefont
  {Berini}(2010)}]{de2010amplification}%
  \BibitemOpen
  \bibfield  {author} {\bibinfo {author} {\bibfnamefont {I.}~\bibnamefont
  {De~Leon}}\ and\ \bibinfo {author} {\bibfnamefont {P.}~\bibnamefont
  {Berini}},\ }\href@noop {} {\bibfield  {journal} {\bibinfo  {journal} {Nature
  Photonics}\ }\textbf {\bibinfo {volume} {4}},\ \bibinfo {pages} {382}
  (\bibinfo {year} {2010})}\BibitemShut {NoStop}%
\bibitem [{\citenamefont {Assawaworrarit}\ \emph {et~al.}(2017)\citenamefont
  {Assawaworrarit}, \citenamefont {Yu},\ and\ \citenamefont
  {Fan}}]{assawaworrarit2017robust}%
  \BibitemOpen
  \bibfield  {author} {\bibinfo {author} {\bibfnamefont {S.}~\bibnamefont
  {Assawaworrarit}}, \bibinfo {author} {\bibfnamefont {X.}~\bibnamefont {Yu}},
  \ and\ \bibinfo {author} {\bibfnamefont {S.}~\bibnamefont {Fan}},\
  }\href@noop {} {\bibfield  {journal} {\bibinfo  {journal} {Nature}\ }\textbf
  {\bibinfo {volume} {546}},\ \bibinfo {pages} {387} (\bibinfo {year}
  {2017})}\BibitemShut {NoStop}%
\bibitem [{\citenamefont {Chang}\ \emph {et~al.}(2014)\citenamefont {Chang},
  \citenamefont {Jiang}, \citenamefont {Hua}, \citenamefont {Yang},
  \citenamefont {Wen}, \citenamefont {Jiang}, \citenamefont {Li}, \citenamefont
  {Wang},\ and\ \citenamefont {Xiao}}]{chang2014parity}%
  \BibitemOpen
  \bibfield  {author} {\bibinfo {author} {\bibfnamefont {L.}~\bibnamefont
  {Chang}}, \bibinfo {author} {\bibfnamefont {X.}~\bibnamefont {Jiang}},
  \bibinfo {author} {\bibfnamefont {S.}~\bibnamefont {Hua}}, \bibinfo {author}
  {\bibfnamefont {C.}~\bibnamefont {Yang}}, \bibinfo {author} {\bibfnamefont
  {J.}~\bibnamefont {Wen}}, \bibinfo {author} {\bibfnamefont {L.}~\bibnamefont
  {Jiang}}, \bibinfo {author} {\bibfnamefont {G.}~\bibnamefont {Li}}, \bibinfo
  {author} {\bibfnamefont {G.}~\bibnamefont {Wang}}, \ and\ \bibinfo {author}
  {\bibfnamefont {M.}~\bibnamefont {Xiao}},\ }\href@noop {} {\bibfield
  {journal} {\bibinfo  {journal} {Nature photonics}\ }\textbf {\bibinfo
  {volume} {8}},\ \bibinfo {pages} {524} (\bibinfo {year} {2014})}\BibitemShut
  {NoStop}%
\bibitem [{\citenamefont {Barton~III}\ \emph {et~al.}(2018)\citenamefont
  {Barton~III}, \citenamefont {Alaeian}, \citenamefont {Lawrence},\ and\
  \citenamefont {Dionne}}]{barton2018broadband}%
  \BibitemOpen
  \bibfield  {author} {\bibinfo {author} {\bibfnamefont {D.~R.}\ \bibnamefont
  {Barton~III}}, \bibinfo {author} {\bibfnamefont {H.}~\bibnamefont {Alaeian}},
  \bibinfo {author} {\bibfnamefont {M.}~\bibnamefont {Lawrence}}, \ and\
  \bibinfo {author} {\bibfnamefont {J.}~\bibnamefont {Dionne}},\ }\href@noop {}
  {\bibfield  {journal} {\bibinfo  {journal} {Physical Review B}\ }\textbf
  {\bibinfo {volume} {97}},\ \bibinfo {pages} {045432} (\bibinfo {year}
  {2018})}\BibitemShut {NoStop}%
\end{thebibliography}%

\end{document}